\documentclass[
]{JFM-FLM_Au}
\usepackage{graphicx}
\usepackage{dcolumn}
\hypersetup{
    colorlinks = true,
    urlcolor   = blue,
    citecolor  = blue, 
}

\usepackage{siunitx}
\usepackage{cleveref}

\crefname{equation}{}{}
\Crefname{equation}{Equation}{Equations}

\crefname{figure}{figure}{figures}
\Crefname{figure}{Figure}{Figures}
\crefname{table}{table}{tables}
\Crefname{table}{Table}{Tables}

\crefname{appendix}{appendix}{appendices}
\Crefname{appendix}{Appendix}{Appendices}

\crefformat{subsection}{\S#2#1#3}
\Crefformat{subsection}{\S#2#1#3}

\crefformat{section}{\S#2#1#3}
\Crefformat{section}{\S#2#1#3}
\crefrangeformat{section}{\S\S#3#1#4 to~#5#2#6}
\Crefrangeformat{section}{\S\S#3#1#4 to~#5#2#6}
\crefmultiformat{section}{\S\S#2#1#3}{ and~#2#1#3}{, #2#1#3}{ and~#2#1#3}
\Crefmultiformat{section}{\S\S#2#1#3}{ and~#2#1#3}{, #2#1#3}{ and~#2#1#3}

\usepackage{tikz}
\usepackage{enumerate}

\renewcommand{\v}[1]{\ensuremath{\boldsymbol{#1}}} 
\let\crossproduct=\times
\renewcommand{\times}{\cdot} 
\newcommand{\gv}[1]{\ensuremath{\mbox{\boldmath$ #1 $}}} 
\newcommand{\diff}{\mathrm{d}}
\renewcommand{\d}[2]{\frac{\diff #1}{\diff #2}} 
\newcommand{\dd}[2]{\frac{\diff^2 #1}{\diff #2^2}} 
\newcommand{\dinl}[2]{{\diff #1}/{\diff #2}}

\newcommand{\pd}[2]{\frac{\partial #1}{\partial #2}}

\newcommand{\grad}[1]{\bnabla #1} 
\renewcommand{\div}[1]{\bnabla \bcdot #1} 
\newcommand{\mean}[1]{\left< #1 \right>}
\newcommand{\laplacian}[1]{\nabla^2 #1}

\newcommand{\pdt}[1]{\partial_t #1}

\newcommand{\Reynolds}{\Rey}
\newcommand{\Prandtl}{\mbox{\textit{Pr}}}  
\newcommand\Peclet{\mbox{\textit{Pe}}}  

\lefttitle{G. Linga, T. Le Borgne and J. Heyman}
\righttitle{Scalar mixing by stagnation points on solid surfaces}

\title{Scalar mixing by stagnation points on solid surfaces}

\author{Gaute Linga\aff{1,2},
 Tanguy Le Borgne\aff{3}
\and Joris Heyman\aff{3}}

\affiliation{
\aff{1}PoreLab, Department of Physics, Norwegian University of Science and Technology, Trondheim, Norway 
\aff{2}PoreLab, The Njord Centre, Departments of Physics and Geosciences, University of Oslo, Oslo, Norway 
\aff{3}Géosciences Rennes, UMR 6118, Université de Rennes, CNRS, Rennes, France 
}

\corresau{Gaute Linga, \email{gaute.linga@ntnu.no}}

\date{\today}
             
\begin{document}
\maketitle

\begin{abstract}
Recent advances have linked mixing in porous media to the deformation of infinitesimal fluid elements.
However, these Lagrangian frameworks generally neglect or simplify interactions with solid surfaces.
We consider transport of thin solute blobs (strips or sheets) near single generic obstacles, in particular along separation lines (separatrices) which connect to stagnation points at solid boundaries.
In this case, we demonstrate that the coupling between flow and diffusion is analytically tractable. 
We provide a complete theory for scalar mixing near upstream stagnation points, which we validate by numerical simulations. 
We identify a mixing time scale $t_B \sim \Peclet ^{1/3}$, where $\Peclet{}$ is the Péclet number, which controls whether scalar decay is driven by fluid stretching or by interaction with the solid boundary.
Asymptotically, for times $t < t_B$, we find that the maximum concentration $c_{\rm max}$ decays as $t^{-5/2}$ and that the width $\sigma$ grows like $t^{1/2}$. For $t > t_B$, we find exponential decay of $c_{\rm max}$ and constant width $\sigma \sim s_B \sim \Peclet^{-1/3}$, where $s_B$ is a 
mixing length scale distinct from that which appears in shear flows. 
The asymptotic scalings of $c_{\rm max}(t)$ and $\sigma(t)$, and their dependence on $\Peclet{}$, are universal to no-slip stagnation points and qualitatively different from stagnation points in open flows.
\end{abstract}



\section{Introduction}
Mixing is at the heart of many transport problems, involving heat or solutes, transported by turbulent or laminar fluid flows \citep{ottino1989kinematics,Villermaux2019,dentz2022mixing}.
Mixing by steady laminar flows is of particular relevance for porous media, where the Reynolds number is generally low \citep{feder2022physics}.
In particular, solute mixing controls dilution and reactions, two processes of key importance for the fate of dissolved elements and contaminants in soils \citep{tartakovsky2013assessment,valocchi2019mixing,rolle2019mixing,sanquer2024microscale,le2026fluid}.
Mixing also drives the heat transfer capacity in exchangers \citep{shah1998heat}, and the selectivity of chromatography columns \citep{liao2015mixing}.
While macrodispersion has received significant attention in porous media \citep{dentz2011mixing,dentz2022mixing}, the dynamics of fluid stretching and mixing at the pore scale have only been investigated recently.

A key feature of pore-scale flow is the no-slip condition, which  enforces that the fluid velocity vanishes at solid boundaries. 
This condition controls the low-velocity tail of the velocity distribution, which has been linked to asymptotic solute dispersion \citep{berkowitz1995characterization,bijeljic2011signature,deanna2017prediction,puyguiraud2021pore}. 
At the pore scale, as a consequence of the heterogeneous velocity field, concentration distributions are known to remain highly heterogeneous over significant time or length scales \citep{Heyman2020}. 
In particular, flow heterogeneity and velocity gradients tend to stretch and sharpen scalar interfaces, increasing scalar gradients and enhancing diffusive fluxes. 

In two-dimensional (2D) pore-scale flow and Darcy flow through heterogeneous scalar permeability fields without discontinuities, material interfaces are stretched at an algebraic rate \citep{LeBorgne2015,dentz2016coupled,lester2021lagrangian,pierce2026pore}.
On the other hand, at the pore scale, steady Stokes flow through three-dimensional (3D) pore networks has been shown to trigger exponential fluid stretching and Lagrangian chaos \citep{Lester2013is,Heyman2020,souzy2020velocity}. 
The latter results in the persistence of scalar gradients over small length scales and underpins the need for a complete description of mixing at the pore scale.
The emergence of chaotic advection has been linked to the omnipresence of stagnation points on the solid boundaries, which organize the flow between attracting and repelling directions and manifolds, producing stretching and folding of material lines in the pore space \citep{turuban2019chaotic,Heyman2020,lester2025unified}.

These observations have been described by the Lagrangian theory of mixing \citep{batchelor1959small,ranz1979applications,ottino1989kinematics,meunier2010diffusive,villermaux2012mixing,martinez2018diffusive,Villermaux2019}, which has recently shown great relevance for capturing incomplete mixing of scalar plumes in porous media \citep{LeBorgne2015,Heyman2020,Heyman2021,izumoto2023enhanced} and in open shear flows \citep{souzy2018mixing}.
This framework describes the action of fluid stretching and molecular diffusion on essentially $(d-1)$-dimensional scalar blobs embedded in $d$-dimensional space.
These blobs are called strips or lamellae ($d=2$) or sheets ($d=3$) and have sharp gradients in the normal direction where compression and diffusion predominantly act \citep{meunier2010diffusive,LeBorgne2015,martinez2018diffusive,Heyman2020,Heyman2021}. 
This approach allows us to simplify 2D and 3D advection-diffusion problems to a one-dimensional (1D) diffusion equation \citep{villermaux2012mixing} with a known solution. 
However, when a blob approaches a solid boundary, the stretching rate varies in the Lagrangian frame, which complicates the lamellar approach. 
Moreover, the solute strip/sheet will diffuse and interact with the solid boundary, qualitatively changing its transverse distribution from the Gaussian it typically assumes in open flows.
How this occurs is generally unknown and motivates this work. 

Hence, despite its importance for fluid-solid reactive processes \citep{steefel2005reactive,aquino2023fluid}, the dynamics of scalar mixing close to solid boundaries has yet to be uncovered.
A series of echo dispersion experiments relying on hydrodynamic reversibility in quasi-2D Stokes flow \citep{oxaal1994irreversible,flekkoy1995hydrodynamic} showed that fluid flow near stagnation points could significantly enhance mixing, measured by orders of magnitude increased effective diffusion. 
Mixing in \emph{saddle flows} -- the flow near stagnation points in the bulk -- is well understood \citep{Villermaux2019,izumoto2023enhanced}.
In this case, the fluid compression rate is constant, which makes the problem particularly amenable to analytical treatment. 
This observation does not hold true close to no-slip stagnation points, where, in fact, the compression rate changes rapidly and vanishes at the boundary.

\begin{figure}
    \centering
    \begin{tikzpicture}[scale=1.0]
    \node [] (A) at (0,0.2cm) {\includegraphics[width=0.35\linewidth]{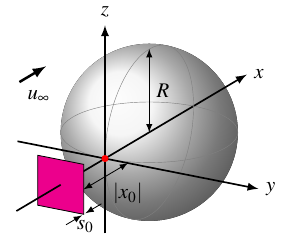}};
    \node [] (B) at (6.6cm,0) {\includegraphics[width=0.64\linewidth,clip,trim=1.6cm 0.45cm 0.65cm 0.4cm]{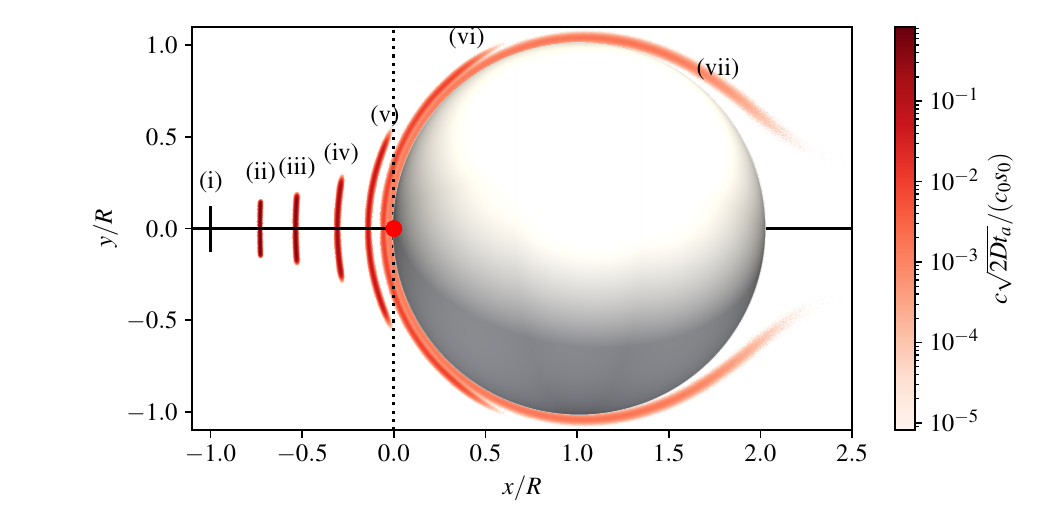}};

    \node [] (Atag) at (-1.8cm,1.95cm) {(\textit{a})};
    \node [] (Btag) at (2.4cm,1.95cm) {(\textit{b})};
    \end{tikzpicture}
    \caption{(a) Schematic illustration of the problem for a spherical obstacle. A thin sheet of dye (show as a magenta-coloured square) is advected by flow towards a solid sphere with a no-slip stagnation point marked by a red circle. The initial thickness of the sheet is $s_0$, the initial position along the mean flow direction is $x_0 < 0$ and the sphere radius is $R$, and the characteristic far-field velocity is $u_\infty$.  (b) Spherical obstacle and scaled scalar concentration $c$ along the cross section $z=0$ at increasing time $t/t_a = $ (i) 0, (ii) 1, (iii) 2. (iv) 4, (v) 8, (vi) 16, (vii) 32. Here, $t_a = R/u_\infty$ is the advective time scale and the concentration scale, $\tilde c = c_0 s_0 / \sqrt{2 D t_a}$, where $D$ is the solute diffusivity, is chosen for consistency with the analytical theory presented in \cref{sec:obstacles_typical}.} 
    \label{fig:sphere}
\end{figure}

The problem we consider is sketched in \cref{fig:sphere}\textit{(a)} for the particular case of a spherical solid obstacle.
Initially, a solute sheet is placed a distance $|x_0|$ from the upstream stagnation point. 
When the flow moves the sheet towards the obstacle, it is acted upon by fluid compression and molecular diffusion. The combination of these processes leads to decay of the maximum concentration, narrowing or broadening of the blob width, and eventually interaction with the solid boundary. 
In \cref{fig:sphere}\textit{(b)}, a cross section of the sheet is shown at different instances of time $t$, as obtained from numerical simulations (see \cref{sec:app:particle_simulation}). 
We focus our analysis on the scalar dynamics near the separatrix, i.e., the black streamline connected to the stagnation point shown as a red dot in both panels of \cref{fig:sphere}.

We develop a quantitative theory of scalar mixing near solid stagnation points which applies to arbitrary velocity profiles along the separatrix. 
We find a two-stage asymptotic behaviour, where, before a mixing time $t_B$, the maximum concentration decays as $t^{-5/2}$ and the blob width grows like $t^{1/2}$. 
After $t_B$, the concentration distribution attains a constant width and its maximum decays exponentially with a rate that depends only on $t_B$.
This behavior is solely driven by the fact that the velocity component normal to the surface depends, to the leading order, quadratically on the distance from the surface near the stagnation point, which is a general feature of no-slip boundaries.
This general result is the main contribution of this work, yielding new insights into the key mechanisms driving scalar decay near solid interfaces that are critical to porous media flows. 
Hence, the work opens possibilities for modeling heterogeneous reaction kinetics in flows through porous media \citep{aquino2023fluid,le2026fluid}.

The paper is organised as follows.
In \cref{sec:infplane}, we introduce the theory for the exactly solvable limit of creeping flow near an infinite plate, a tractable limit that isolates the essential physics.
In \cref{sec:general}, we generalize the theory to arbitrary velocity profiles, obtaining expressions for the maximum scalar concentration and distribution width that together characterize mixing.
In \cref{sec:obstacles_typical} we apply the theory to creeping flow past a sphere, for which the velocity field is known analytically and explicit solutions can be obtained.
This setting is particularly relevant to granular porous media.
Finally, in \cref{sec:discussion}, we discuss implications of our results in light of established literature, including on reactive transport.


\section{Mixing by flow close to an infinite plate: An exactly soluble case}
\label{sec:infplane}



\begin{figure}
    \centering
    \includegraphics[width=0.6\linewidth]{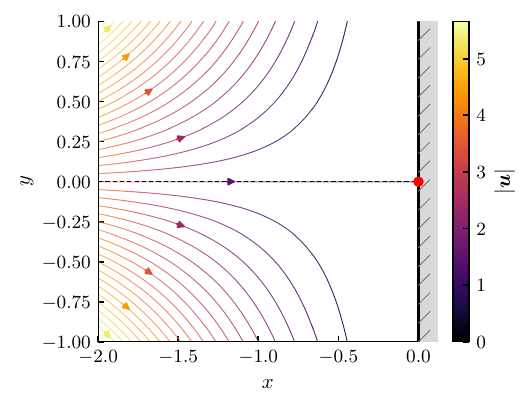}
    \caption{Streamlines of flow close to an infinite plate. The flow domain is the infinite half-space $x<0$, the separatrix is at $y=0$ and the stagnation point is indicated by a red circle at $x=y=0$. The velocity field $\v u$ corresponds to \cref{eq:u_infplate_3d} with $z=0$ and $\alpha=\beta=1$.}
    \label{fig:infplate_streamlines}
\end{figure}

As an introductory case, we consider a solute strip advected by steady, incompressible flow toward an infinite-plane solid boundary located at $x=0$ with a mean flow along $x$ (see \cref{fig:infplate_streamlines}).
This idealized velocity field can be derived, for example, by Taylor expansion of analytical velocity fields close to the stagnation point located at the origin (see \cref{sec:app:sphere}), and is often referred to as `stagnation point flow' \cite[p.~285]{batchelor2000introduction}.
Taking the fluid domain to be the three-dimensional infinite half-space $x \leq 0$, these velocity fields can in general be written on the form
\begin{align}
    \v u = \begin{bmatrix}
        \alpha x^2 \\
        - 2 \alpha \beta x y \\
        - 2 \alpha (1 - \beta) x z
    \end{bmatrix},
    \label{eq:u_infplate_3d}
\end{align}
where $\alpha > 0, \beta \in [0, 1]$ are problem-dependent constants.
It is easy to verify that \cref{eq:u_infplate_3d} satisfies the no-slip condition, $\v u = \v 0$ at the solid boundary $x = 0$, and the incompressibility condition $\div \v u = 0$.
Given the pressure field $p = p_\infty + 2 \alpha \mu x$, where $p_\infty$ is some reference pressure, it also satisfies the momentum balance from the Stokes equations,
\begin{align}
    \mu \laplacian \v u = \grad p,
\end{align}
where $\mu$ is the dynamic viscosity.
This equation should be valid sufficiently close to a no-slip boundary, as the Reynolds number (given by $\Reynolds = \rho U R / \mu$ where $\rho$ is the fluid density, $R$ and $U$ are typical length and velocity scales, respectively) is sufficiently small for inertial terms to be neglected.
Since the domain in this case does not have an intrinsic length scale, the flow compression is fully described by the parameter $\alpha$ that characterizes the growth of the compression rate away from the cylinder.
For simplicity, in the following we will consider the 2D flow field where $\beta = 1$, i.e.\ where all fluid deformation is in the $xy$ plane.

The transport of a passive scalar concentration $c(\v x, t)$ in space $\v x$ and time $t$ is governed by the advection-diffusion equation:
\begin{equation}
    \pdt c + \v u \cdot \grad c = D \laplacian c,
    \label{eq:ade3d}
\end{equation}
with the no-flux boundary condition $\hat{\v n} \cdot \grad c = 0$ imposed at the fluid-solid interface $x=0$. 
Here, $D$ is the molecular diffusivity and $\hat{\v n} = \hat{\v e}_x$ is the surface normal.
The initial condition is taken to be a Gaussian distribution, $c(\v x, 0) = c_{\rm init}(x)$ where
\begin{align}
    c_{\rm init} (x) = c_{0} \exp\left( - \frac{(x-x_0)^2}{2 s_0^2} \right).
    \label{eq:c_0_x}
\end{align}
The initial concentration distribution is characterized by the width $s_0$, the maximum concentration $c_{0}$, and the initial position of the line $x_0 < 0$ (meaning that the initial distance is $|x_0| = -x_0$).
Because the initial distribution and the $x$-component of the velocity field depend only on $x$, the problem remains one-dimensional. This implies that the concentration evolution equation \cref{eq:ade3d} simplifies to
\begin{align}
    \pdt c = \alpha x^2 \partial_x c + D \partial_x^2 c,
    \label{eq:c_evo_1d}
\end{align}
using \cref{eq:u_infplate_3d}. From the two parameters $\alpha, D$ we deduce an intrinsic length scale $s_B$ and a time scale $t_B$: 
\begin{equation}
	s_B = (D/\alpha)^{1/3}
	\quad \textrm{and} \quad
	t_B = D^{-1/3}\alpha^{-2/3}.
	\label{eq:stB_def}
\end{equation}
We identify $s_B$ as the Batchelor scale, which is proportional to the equilibrium thickness of a strip where diffusion and compression balance \citep{batchelor1959small,villermaux2012mixing}.
Correspondingly, we identify $t_B$ as the mixing time, proportional to the time required for the strip to reach this thickness \citep{Villermaux2019}.

In addition to these scales that are intrinsic to \cref{eq:c_evo_1d}, the initial conditions will play a role through the initial distance $|x_0|$ and the initial thickness $s_0$.
In particular, the time scale $t_0 = 1/(-x_0 \alpha)$ (the advective time scale based on the initial velocity) will be a key to the theoretical description of the concentration evolution.

\subsection{Numerical study of the solute strip evolution}

\begin{figure}
    \centering
    \includegraphics[width=0.7\textwidth]{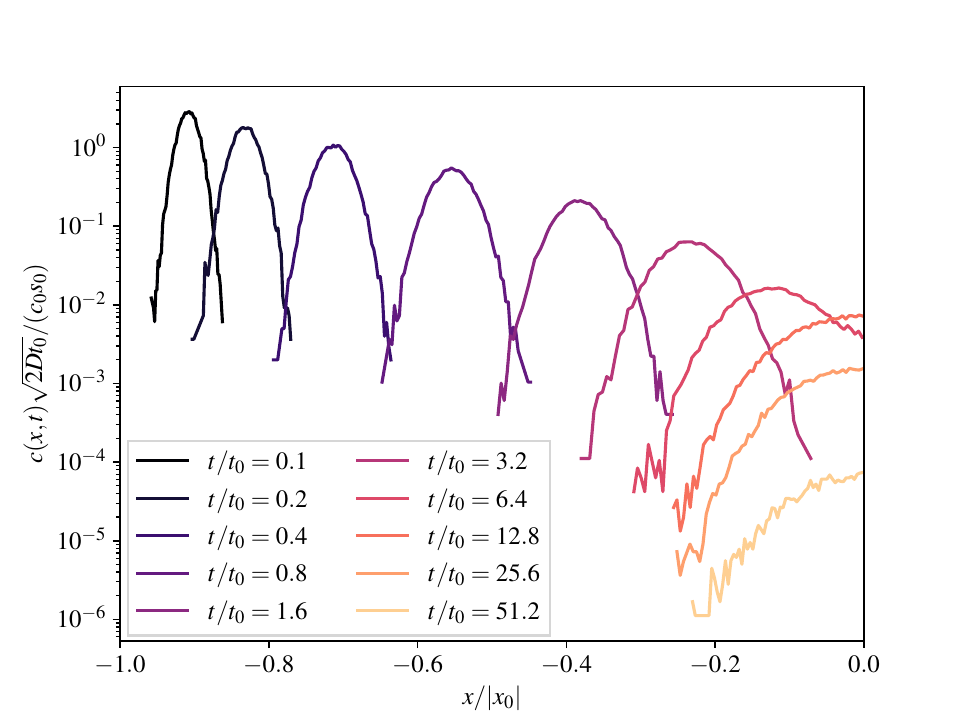}
    \caption{The evolution of the concentration of a solute strip along the separatrix in the case of flow toward an infinite plate at $x=0$, where the velocity field is given by \cref{eq:u_infplate_3d} with $\alpha=\beta=1$.
    The initial distribution is a delta-peak at $x = x_0 = -1$, and the molecular diffusion coefficient is set to $D=10^{-3}$.
    }
    \label{fig:c_infplate}
\end{figure}

To investigate the mixing of a strip of solute in the velocity field $\v u$ defined in \cref{eq:u_infplate_3d}, we solve \cref{eq:ade3d} using a particle tracking method outlined in \cref{sec:app:particle_simulation}.
We use $N = 10^6$ advected diffusive particles, and the flow parameters $\alpha = \beta = 1$.
The initial state corresponds to a thin Gaussian strip \eqref{eq:c_0_x} with half-width $s_0 \to 0$, positioned at $x_0 = -1$ and defined in a region $y \in [-\ell/2, \ell/2]$.

The evolution of the concentration field (along the separatrix) in time is shown in \cref{fig:c_infplate}. 
The concentration scale $\tilde c = c_0 s_0 /\sqrt{2 D t_0} $ is chosen to normalize the concentration in a way that is well-defined for an initially infinitely thin strip (as $s_0 \to 0$, we have that $c_0 \to \infty$ while $c_0 s_0$ is finite) and for compatibility with the theory presented in \cref{sec:theory_infplate}.
We observe that the strip is transported by the flow field with a near-Gaussian shape, widens, slows down, and is gradually compressed towards the solid boundary, achieving a short-tailed one-sided shape.
The overall concentration generally decreases.

For a more quantitative look, we will systematically investigate the time evolution of two key quantities that characterize the concentration distribution $c(x, t)$:
\begin{itemize}
    \item The distribution half-width, $\sigma (t)$, related to the variance $\sigma^2 = \mean{x^2} - \mean{x}^2$,
    \item The maximum concentration $c_{\rm max} (t)$.
\end{itemize}
These two quantities represent essential information on mixing near the stagnation point.
From them, we can deduce two other key quantities for mixing:
The product $\sigma c_{\rm max}$ will be unchanged by diffusion in the direction normal to the strip/sheet, i.e.\ in the absence of strong diffusion along the strip, it should only be influenced by compression and thus reflects a local mass conservation.
The ratio $c_{\rm max} / \sigma $ is a basic measure of the magnitude of concentration gradients, the ultimate driver of mixing.
In addition, we will investigate the mean position of the strip, $\mean{x}$. 
Details of how we compute the mentioned time-dependent quantities are given in \cref{sec:app:particle_simulation}.

\begin{figure}
    \centering
    \includegraphics[width=\textwidth]{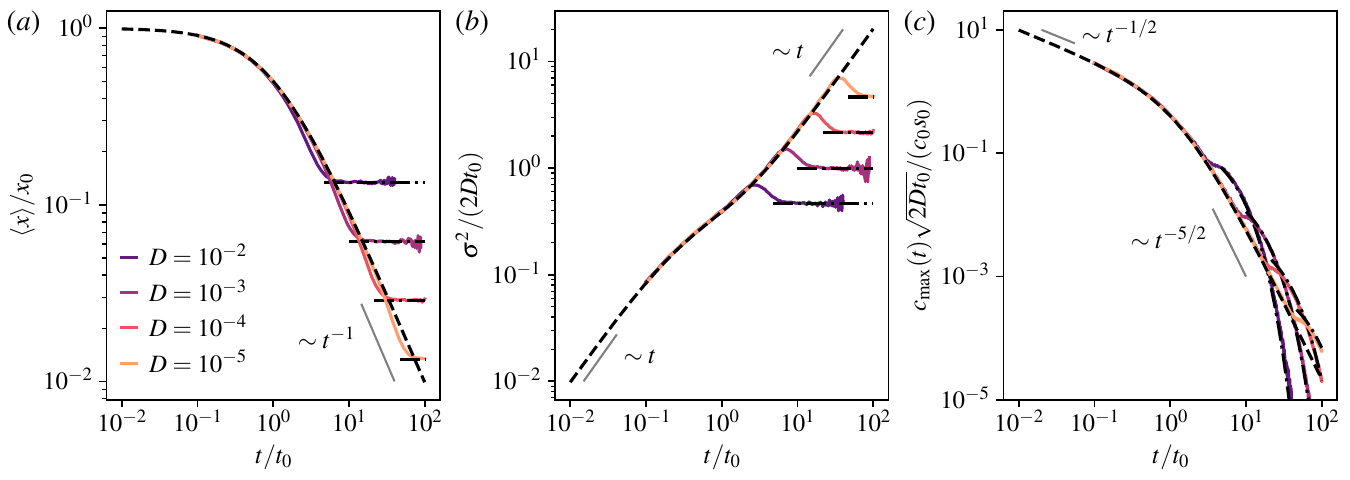}
    \caption{Evolution in time $t$ of three key mixing measures of solute strips with different diffusion constants $D$ approaching a no-slip stagnation point at an infinite plate:
    (\textit{a}) Mean position of the strip,
    (\textit{b}) positional variance (i.e.\ squared half-width), and 
    (\textit{c}) maximum concentration.
    The fluid velocity field is given by \cref{eq:u_infplate_3d} with $\alpha=\beta=1$.
    In all panels, the colored solid lines represent numerical simulations and the dashed black lines correspond to the theory in \cref{sec:theory_infplate}.
    The dash-dotted black lines correspond to the saturation regime described in \cref{sec:infplane_sat}.
    }
    \label{fig:infplate_panels}
\end{figure}

In \cref{fig:infplate_panels}, we plot the mean position, half-width, and maximum concentration of $c (x, t)$ as a function of scaled time $t/t_0$ for a range of different molecular diffusion coefficients $D \in [10^{-5}, 10^{-2}]$.
We fix the initial position $x_0 = -1$ and impose an infinitely thin initial strip width, $s_0 \to 0$.
The mean position normalized by the initial position, $\mean{x} / x_0$, follows the same curve towards the origin regardless of the initial position (\cref{fig:infplate_panels}(\textit{a})), until a certain point where the position saturates at a value of $x = x_*$ depending on $D$. 
A similar behaviour is seen in \cref{fig:infplate_panels}(\textit{b}) for the measurements of $\sigma^2 / 2 D t_0$ which initially follow a universal curve before saturating at a similar time as $x(t)$.
In particular, $\sigma$ increases with time, initially purely diffusively ($\sigma^2 \sim 2 D t$, as expected for an initially sharp distribution), then at a slower relative rate, and finally diffusively ($\sigma^2 \propto t$) again, before plateauing at a level dependent on the parameters $D, \alpha$.
The peak concentration decays, first algebraically and eventually exponentially (\cref{fig:infplate_panels}(\textit{c})). 
The onset times of the exponential decay coincide with the plateauing of $\mean{x}$ and $\sigma$.

\begin{figure}
    \centering
    \includegraphics[width=\textwidth]{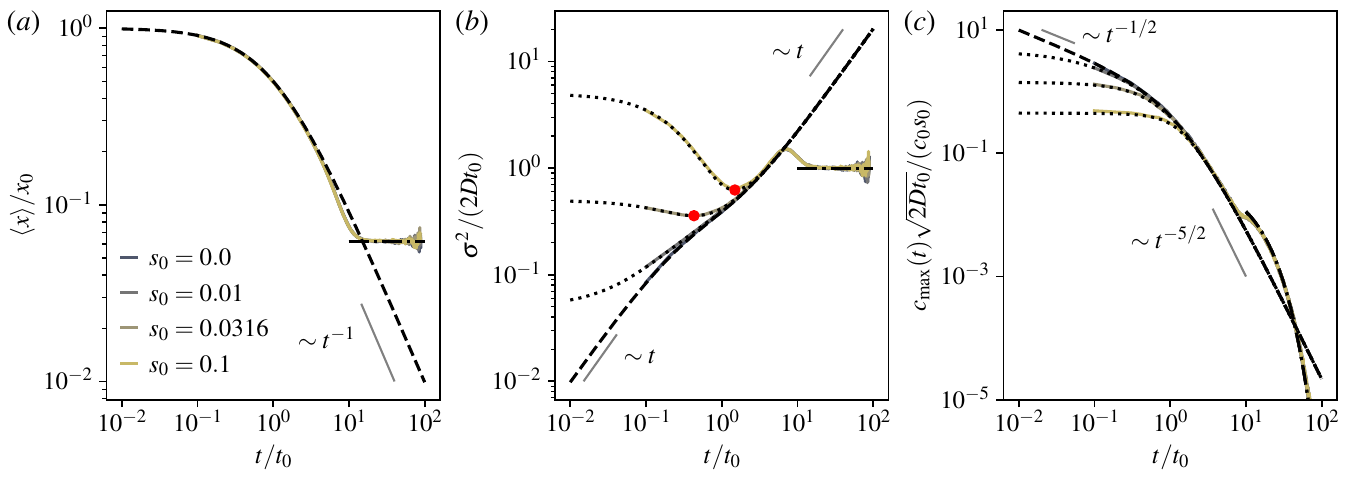}
    \caption{Temporal evolution of solute strips of different initial thickness $s_0$ approaching an infinite plate for a fixed diffusion coefficient $D=10^{-3}$.
    (\textit{a}) The mean position of the strip is unaffected by $s_0$. (\textit{b}) The positional variance depends on strongly on the initial thickness and exhibits compression if $s_0$ is large enough, leading to a second mixing scale $s_m$ and time $t_m$ indicated by a red circle. (\textit{c}) The maximum concentration across the strip can go directly to the sharp algebraic $t^{-5/2}$ scalar decay for sufficiently high $s_0$.
    The fluid velocity field is the same as in \cref{fig:infplate_panels} and given by \cref{eq:u_infplate_3d} with $\alpha = \beta = 1$.
    The colored solid lines represent numerical simulations and the dashed black lines corresponds to the analytical theory in \cref{sec:theory_infplate}.
    The dash-dotted black lines correspond to the saturation regime described in \cref{sec:infplane_sat}.
    }
    \label{fig:infplate_panels_s0}
\end{figure}

It is also of interest to investigate the influence of a finite initial thickness $s_0 > 0$. 
In \cref{fig:infplate_panels_s0}, we plot the same quantities as in \cref{fig:infplate_panels}, but with the diffusion coefficient fixed at $D=10^{-3}$ and with different $s_0 \in [10^{-2}, 10^{-1}]$ in addition to the baseline $s_0 \to 0$ case.
Although there is no visible difference in behaviour for the mean position of the strip (\cref{fig:infplate_panels_s0}(\textit{a}), the half-width $\sigma$ obeys a markedly different behaviour. 
Due to the initial condition $\sigma(0) = s_0$, it starts out higher than the $s_0 \to 0$ curve, but then gradually approaches it as $t \sim t_0$. 
For sufficiently high $s_0$, the strip is compressed for a period and reaches a minimum which we denote $(t_*, s_*)$. 
This minimum gives rise to a mixing time and mixing scale distinct from previously identified $(t_B, s_B)$, which depends on $s_0$ is analogous to how the mixing time and length scales are usually defined in shear flows \citep{souzy2018mixing}.
In \cref{fig:infplate_panels_s0}(\textit{c}) we show such minima as red dots for different $s_0$. 
Using the point where $\sigma$ attains a minimum 
The maximum concentration is shown in \cref{fig:infplate_panels_s0}(\textit{c}) and starts as lower than the $s_0 \to 0$ case. 
This is a consequence of the chosen concentration scale $\tilde c$, which is proportional to the total mass across the profile, i.e.\ $\propto c_0 s_0$. 
The maximum concentration remains constant approximately until $t \sim t_*$, and then approaches the universal $s_0 \to 0$ curve.
The development beyond this is point is indistinguishable from the $s_0 \to 0$ case shown in \cref{fig:infplate_panels}.

\subsection{Theory for the infinite plate}
\label{sec:theory_infplate}
We now seek to explain the evolution of the strip theoretically. 
To do so, we employ the lamellar theory, a framework for relating fluid stretching to the evolution of solute blobs \citep{ranz1979applications,ottino1989kinematics,LeBorgne2015,Villermaux2019}.
The key idea is to map the problem from the Eulerian description \cref{eq:ade3d} to a Lagrangian frame co-moving with a material line or surface, reconstructing the concentration distributions via the stretching history of individual fluid lamellae.
We briefly recall the necessary parts of this well-established theoretical framework in \cref{sec:app:lamella}.

For a strip with a normal direction along $x$, the lamellar theory \citep{ranz1979applications,Villermaux2019} predicts the Gaussian concentration distribution around it as
\begin{align}
    c (x, t) = c_{\rm max}(t) \exp\left( - \frac{(x-\bar x(t) )^2}{2\sigma^2(t)} \right).
    \label{eq:gaussianc_1}
\end{align}
The three time-dependent quantities $\bar x(t)$, $\sigma(t)$ and $c_{\rm max}(t)$ fully characterize the distribution and are the same as plotted in \cref{fig:infplate_panels}.
For convenience of notation, we will skip the overbar on $\bar x$ in the following, since we are primarily interested in the evolution of these three quantities.

The position of the material line $x(t)$, i.e.\ the approximate mean position of the strip, is described by
\begin{align}
    \d x t = u_x = \alpha x^2,
\end{align}
which can be integrated to yield
\begin{align}
    x (t) = x_0 \left[1 + \frac{t}{t_0} \right]^{-1}.
    \label{eq:x1d}
\end{align}
This expression is compared with the simulated data in \cref{fig:infplate_panels}(\textit{a}) and matches well until the numerical data saturate (which happens at a time $t \sim t_B$, as we shall see in the forthcoming \cref{sec:infplane_sat}). 
The same behaviour is seen \cref{fig:infplate_panels_s0}(\textit{a}), where the finite initial thickness has no significant influence on the mean position $x$, and theoretical expression matches well until the saturation plateau is reached.

The half-width (i.e.\ spatial standard deviation) of the distribution is given by the lamellar theory \citep{Villermaux2019} (summarized in \cref{sec:app:lamella}) as 
\begin{equation}
    \sigma(t) = s(t) \sqrt{1 + \frac{2 D}{s_0^2} \tau(t)} ,
    \label{eq:sigmadef}
\end{equation}
where $s(t)$ is the \emph{striation thickness}, i.e.\ the equivalent width of the distribution in the absence of diffusion, where by construction $s(0) = s_0$.
In \cref{eq:sigmadef}, we have defined the \emph{warped time} $\tau$ as
\begin{align}
    \tau = s_0^2 \int_0^t \frac{\diff t'}{s^2 (t')}.
    \label{eq:taudef}
\end{align}
In order to obtain a closed-form expression for $\sigma(t)$ using \cref{eq:sigmadef}, we need an expression for $s(t)$ and thereby $\tau(t)$ through \cref{eq:taudef}. 
We note that the compression rate $\gamma$ can be expressed in two ways:
\begin{equation}
    \gamma = -\frac{1}{s} \d{s}{t} 
    \quad \textrm{and} \quad
    \gamma = -\pd{u}{x} = -2 \alpha x = \frac{2}{t_0} \left[ 1 + \frac{t}{t_0} \right]^{-1},
    \label{eq:compress_00}
\end{equation}
where we have used \cref{eq:x1d} in the last equality. Equating the two expressions for $\gamma$, we may solve for the striation thickness to get
\begin{align}
    s(t) = s_0 \left[ 1 + \frac{t}{t_0} \right]^{-2}.
    \label{eq:st_infplate}
\end{align}
For the warped time, inserting \cref{eq:st_infplate} into \cref{eq:taudef}, the integral becomes
\begin{align}
    \tau(t) = \int_0^t \left[ 1 + \frac{t'}{t_0} \right]^{4} \, \diff t' = \frac{t_0}{5} \left[ \left( 1 + \frac{t}{t_0} \right)^{5} - 1 \right].
    \label{eq:tau1d}
\end{align}
Insertion of the latter \cref{eq:tau1d} into \cref{eq:sigmadef} yields the following relatively simple expression for the half-width:
\begin{align}
 \sigma (t) = \left[ 1 +\frac{t}{t_0} \right]^{-2} \sqrt{s_0^2 + \frac{2 D t_0}{5} \left[ 
    \left( 1 + \frac{t}{t_0} \right)^{5} - 1 \right] } ,
    \label{eq:sigma_infplate}
\end{align}
which in the limit $s_0 \to 0$ becomes even simpler:
\begin{align}
    \sigma (t) = \left[ 1 +\frac{t}{t_0} \right]^{-2} \sqrt{\frac{2 D t_0}{5} \left[ 
    \left( 1 + \frac{t}{t_0} \right)^{5} - 1 \right] }
    .
    \label{eq:sigma_infplate_simpl}
\end{align}
As seen in \cref{fig:infplate_panels}(\textit{b}), where \cref{eq:sigma_infplate_simpl} is plotted as a dashed black line, the theory provides excellent agreement with the simulation data until the saturation plateau is reached.

The complete expression \eqref{eq:sigma_infplate} for the half-width $\sigma$ also allows us to check the case for finite initial thickness $s_0$. 
As shown in \cref{fig:infplate_panels_s0}(\textit{b}), the agreement between \cref{eq:sigma_infplate} and the simulated data for different diffusion coefficients $D$ is excellent.
Moreover, using \cref{eq:sigma_infplate}, we can find explicit expressions for the intermediate mixing scales $(t_m, s_m)$.
Differentiating \cref{eq:sigma_infplate} with respect to time $t$ and equating to zero, we find that a local minimum of $\sigma$ is achieved when $t = t_m$ where
\begin{align}
    \frac{t_m}{t_0} = \left[ 20 \left( \frac{s_0^2}{2Dt_0} - \frac{1}{5} \right) \right]^{1/5} - 1.
    \label{eq:tstar_infplate}
\end{align}
This minimum exists only when $s_0 > \sqrt{Dt_0/2}$ and yields the length scale
\begin{align}
    \frac{s_m}{\sqrt{2 D t_0}} = \frac{1}{2}\left[ 20 \left( \frac{s_0^2}{2Dt_0} - \frac{1}{5} \right) \right]^{1/10} .
    \label{eq:sstar_infplate}
\end{align}
We therefore see that for a finite initial thickness $s_0 \gg \sqrt{2 D t_0}$, we identify the scalings $t_m \sim s_0^{2/5}$ and $s_m \sim s_0^{1/5}$. 
The exponents are distinct from those found in shear flows, where the corresponding $t_m \sim s_0^{2/3}$ and $s_m \sim s_0^{1/3}$ \citep{souzy2018mixing}. 
The superimposed red dots indicating the minima of $\sigma$ in \cref{fig:infplate_panels_s0} are given by \cref{eq:tstar_infplate,eq:sstar_infplate} and agree well with the simulations.

The maximum concentration in \cref{eq:gaussianc_1} is obtained from the lamellar theory \citep{ranz1979applications,Villermaux2019} as summarized in \cref{sec:app:lamella}:
\begin{equation}
    c_{\rm max}(t) = \frac{c_0}{\sqrt{1 + \frac{2 D}{s_0^2} \tau(t)}} .
    \label{eq:cmaxasft}
\end{equation}
Using \cref{eq:tau1d}, \cref{eq:cmaxasft} becomes
\begin{align}
    c_{\rm max}(t) &= \frac{c_0 s_0}{\sqrt{s_0^2 + \frac{2 D t_0}{5} \left[ \left( 1 + \frac{t}{t_0} \right)^{5} - 1 \right]}}
    = \frac{c_0 s_0}{\sqrt{2 D t_0}} \mathscr{C}\left( \frac{t}{t_0}; \frac{s_0}{\sqrt{2Dt_0}} \right).
    \label{eq:cmax1d}
\end{align}
Here, we have defined the scaling function
\begin{align}
    \mathscr{C}(\tilde t; \tilde s_0) = \left[\tilde s_0^2 + \frac{1}{5} \left[ \left( 1 + \tilde{t} \right)^{5} - 1 \right] \right]^{-1/2}.
\end{align}
In the limit of $s_0 \to 0$, \cref{eq:cmax1d} simplifies to 
\begin{align}
    c_{\rm max}(t) \simeq \frac{c_0 s_0 \sqrt{5} / \sqrt{2 D t_0}}{\sqrt{ \left( 1 + \frac{t}{t_0} \right)^{5} - 1 }}.
    \label{eq:cmax1d_simp}
\end{align}
This expression provides a proper theoretical justification for the normalization factor $\tilde c = c_0 s_0 / \sqrt{2 D t_0} $ used previously.
In particular, all simulation data of $c_{\rm max} / \tilde c$ are predicted to follow the same curve given by \cref{eq:cmax1d} (or \cref{eq:cmax1d_simp} in the case of $s_0 \to 0$).

In \cref{fig:infplate_panels}(\textit{c}) we confirm this for $s_0$, where the theoretical expression \eqref{eq:cmax1d_simp} is shown as a dashed black line superimposed on the simulation data.
Consistently with the trend for $x$ and $\sigma$, the theoretical expression predicts the simulation data up to the point where the length scale plateaus, and the concentration starts to decay exponentially.
Similarly, for finite $s_0$, as shown in \cref{fig:infplate_panels_s0}(\textit{c}), the theoretical expression \eqref{eq:cmax1d} accurately predicts the temporal evolution of the maximum concentration until the point of saturation.

\subsection{Saturation and breakdown of the lamellar theory}
\label{sec:infplane_sat}
Having obtained a quantitative theory for the evolution of the strip within the lamellar theory, we now address the saturation regime. In particular, the assumption of a Gaussian distribution, which is central to the lamellar theory, breaks down when the width of the distribution becomes comparable to the distance to the wall.
This occurs at a time $t_*$ when $\sigma (t_*)\sim |x(t_*)|$. 
Using \cref{eq:x1d,eq:sigma_infplate} and denoting $x_* = x(t_*)$, we obtain
\begin{align}
    \sigma^2(t_*) = \left( 
        \frac{x_*}{x_0} 
    \right)^4 \left\{ s_0^2 + \frac{2Dt_0}{5} \left[ \left( \frac{x_*}{x_0}\right)^{-5} - 1 \right] \right\}
    \sim x_*^2
\end{align}
In the limit $x_*/x_0 \ll 1$ we can solve for $x_*$ to obtain:
\begin{align}
    - x_*  
    \sim \sqrt[3]{\frac{2 D }{5 \alpha}} \sim s_B ,
\end{align}
where we have made use of the definition of $s_B$ in \cref{eq:stB_def}.
We can solve for the time $t_*$ by again using \cref{eq:x1d}:
\begin{align}
    t_* = t_0 \left[ \frac{x_0}{x_*} - 1\right] \sim \sqrt[3]{\frac{5}{2D\alpha^2}} \sim t_B,
\end{align}
again using \cref{eq:stB_def} for the definition of $t_B$ in the last step.

From this point on, the distribution is no longer Gaussian, but becomes squeezed towards the solid boundary with a steep (superexponential) tail for large distances $|x|$, which can be seen at large times, $t > 10 t_0$, in \cref{fig:c_infplate}.
Asymptotically with time, the distribution approaches a constant shape and where only the magnitude $c_{\rm max}(t)$ decays exponentially with time.
We show in \cref{sec:app:finiteD} that the time-asymptotic solution ($t \gg t_B$) is given by
\begin{align}
    c(x, t) \sim c_0 \exp\left( - \frac{\lambda_1 t}{t_B} \right) \phi_1\left( \frac{|x|}{s_B} \right),
    \label{eq:c_sat}
\end{align}
where $\lambda_1$ is the smallest positive eigenvalue and $\phi_1(z)$ is the corresponding eigenfunction solving the problem $ x^2 \phi'' (x) + \phi'(x) + \lambda \phi = 0, \phi'(0) = 0$. 
The superexponential spatial tail is given by the $z \gg 1$ asymptote $\phi_1(z) \sim z^{-2} \exp(-z^3/3)$, which is seen to fall off more rapidly than a Gaussian.
We numerically determine $\lambda_1 \simeq 1.241$ and the moments of $\phi_1$, which yield
\begin{subequations}
\begin{align}
    x (t\to\infty) &\simeq 0.620 s_B, \label{eq:x_sat}\\
    \sigma (t\to\infty) &\simeq 0.447 s_B.
    \label{eq:sigma_sat}
\end{align}\label{eq:scales_asymptotic_infplate}
\end{subequations}
These scales are superimposed as horizontal dash-dotted lines in \cref{fig:infplate_panels}(\textit{a}) and (\textit{b}), respectively, and show excellent agreement with the numerical simulation data for different diffusion coefficients $D$.
For completeness, we also show the same predictions in \cref{fig:infplate_panels_s0}(\textit{a}) and (\textit{b}), respectively, for different $s_0$ which all approach the same value.
 
From \cref{eq:c_sat} we can also read off the maximum concentration as
\begin{align}
    c_{\rm max} (t) \sim \exp\left( - \frac{\lambda_1 t}{t_B} \right)
    \quad \textrm{for}\quad 
    t \gg t_B.
    \label{eq:cmax1dsat}
\end{align}
In \cref{fig:infplate_panels}(\textit{c}), we compare with numerical simulations, which agree well with the prediction shown by a dash-dotted black line.

Note that the prefactor in \eqref{eq:cmax1dsat} is not determined analytically yet; we do so in the next section.
%

\subsection{Unified description and data collapse}
\label{sec:unified_infplate}
\begin{figure}
    \centering
    \includegraphics[width=0.8\textwidth]{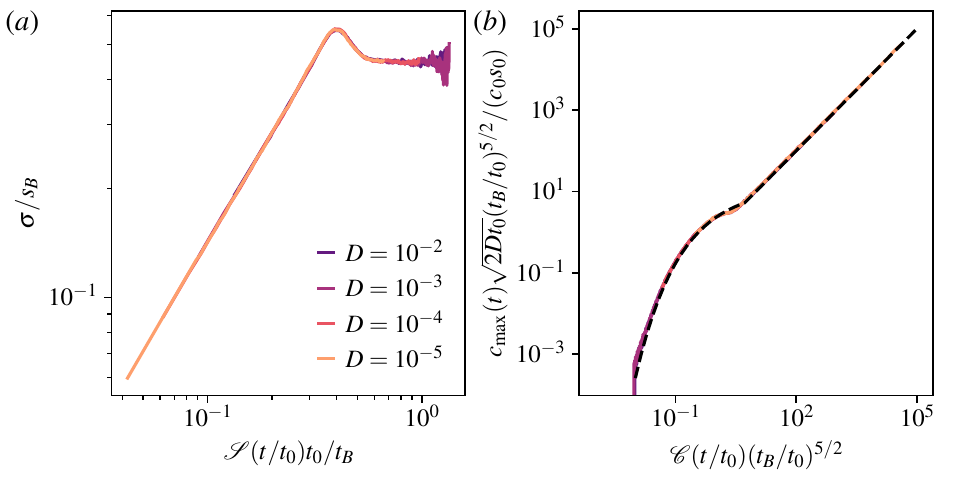}
    \caption{Data collapse of the same data as presented in \cref{fig:infplate_panels}. 
    The colored solid lines represent simulations with different diffusion coefficients $D$.
    (\textit{a}) The simulated data for the half-width $\sigma$ normalized by the Batchelor scale, $s_B$, collapses onto a single curve when plotted as a function of the single composite variable $\mathscr{S}(t/t_0; s_0/\sqrt{2 D t_0}) \sqrt{ t_0 / t_B} $, as predicted by \cref{eq:unified_Sigma_infplate}. 
    (\textit{b}) As predicted by \cref{eq:cmax_unified_infplane}, the simulated data for the maximum concentration $c_{\rm max}$, multiplied by the factor $\sqrt{2Dt_0} (t_B/t_0)^{5/2}/(c_0 s_0)$, collapses onto a single curve when plotted as a function of $\xi$ (\cref{eq:xidef}). 
    By superimposing the function $H(\xi)$ shown as a black dashed line, defined in the right hand side of \cref{eq:cmax_infplane_4}, we see that this universal curve is very close to $H(\xi)$.
    }
    \label{fig:infplate_collapse}
\end{figure}

We now have analytical expressions for the scalar decay both well before and after the strip reaches the solid boundary, which occurs at time $t \sim t_B$. 
For the half-width of the strip, we have from \cref{eq:sigma_infplate_simpl,eq:sigma_sat} that
\begin{align}
    \sigma = \begin{cases}
    \sqrt{2 D t_0} \mathscr{S} \left( \frac{t}{t_0}; \frac{s_0}{\sqrt{2Dt_0}}\right)
    &\quad \textrm{for} \quad t \ll t_B, \\
    0.447 s_B &\quad \textrm{for} \quad t \gg t_B.
    \end{cases}
\end{align}
Dividing by $s_B$, we find
\begin{align}
    \frac{\sigma}{s_B} = \begin{cases}
    \sqrt{2} \left( \frac{t_0}{t_B}\right)^{1/2} \mathscr{S} \left( \frac{t}{t_0}; \frac{s_0}{\sqrt{2Dt_0}}\right)
    &\quad \textrm{for} \quad t \ll t_B, \\
    0.447 &\quad \textrm{for} \quad t \gg t_B.
    \end{cases}
\end{align}
This expression suggests that the scaled width $\sigma / s_B$ depends only on the scaled variable $\left( {t_0} / {t_B}\right)^{1/2} \mathscr{S} \left( {t} / {t_0}; {s_0} / {\sqrt{2Dt_0}}\right)$ over the entire data set, independently of the specific parameters ($\alpha, D, x_0, s_0$).
We plot in \cref{fig:infplate_collapse}(\textit{a}) this relation 
based on our simulation data shown in \cref{fig:infplate_panels}(\textit{b}).
We find a remarkably good data collapse for all simulations (where $s_0 = 0$).
This observation means that the width is well described by
\begin{align}
    \sigma = s_B \, G\left( \sqrt{\frac{t_0}{t_B}} \mathscr{S} \left( \frac{t}{t_0} ; \frac{s_0}{\sqrt{2Dt_0}} \right) \right) ,
    \label{eq:unified_Sigma_infplate}
\end{align}
where $G(z)$ is a universal scaling function with the properties $G(z) \to \sqrt{2} z$ for $z \ll 1$ and $G(z) \to \textrm{const.} \simeq 0.447$ for $z \gg 1$.

For the maximum concentration, on the other hand, we have
\begin{align}
    c_{\rm max} = \begin{cases}
    \frac{c_0 s_0} {\sqrt{2 D t_0}} \mathscr{C} \left( \frac{t}{t_0}; \frac{s_0}{\sqrt{2Dt_0}}\right)
    &\quad \textrm{for} \quad t \ll t_B, \\
    K e^{-\lambda_1 t / t_B} &\quad \textrm{for} \quad t \gg t_B,
    \end{cases}
\end{align}
where we have yet to determine the time-independent prefactor $K$. 
To do so, we simplify and extrapolate the two branches to an intermediate crossover time $t_\crossproduct \sim t_B$: 
\begin{align}
    \frac{c_{\rm max} {\sqrt{2 D t_0}} }{ c_0 s_0 } = \begin{cases}
     \mathscr{C} \left( \frac{t}{t_0}; \frac{s_0}{\sqrt{2Dt_0}}\right)
    &\quad \textrm{for} \quad t \leq t_\crossproduct, \\
    \frac{\sqrt{2 D t_0}}{c_0 s_0} K e^{-\lambda_1 t / t_B} &\quad \textrm{for} \quad t > t_\crossproduct.
    \end{cases}
    \label{eq:cmax_infplane_2}
\end{align}
Here, we have also made both sides dimensionless by dividing by the concentration scale $\tilde c$. 
To connect the two branches, we must have
\begin{align}
    K = \frac{c_0 s_0} {\sqrt{2 D t_0}} \mathscr{C} \left( \frac{t_\crossproduct}{t_0}; \frac{s_0}{\sqrt{2Dt_0}}\right) e^{\lambda_1 t_\crossproduct / t_B}
    \simeq \frac{c_0 s_0} {\sqrt{2 D t_0}} \sqrt{5} \left( \frac{t_0}{t_B} \right)^{5/2} \chi^{-5/2} e^{\lambda_1 \chi},
\end{align}
where we have used the definition $\chi = t_\crossproduct/t_B$, the approximation $t_\crossproduct \gg t_0$, and
\begin{align}
    \mathscr{C} \left( \tilde t; \tilde s_0 \right) \simeq \sqrt{5} \tilde{t}^{-5/2} \quad \textrm{for} \quad \tilde t \gg 1. 
    \label{eq:scrC_approx}
\end{align}
Multiplying \cref{eq:cmax_infplane_2} by $(t_B/t_0)^{5/2}$, we obtain
\begin{align}
    \frac{c_{\rm max} {\sqrt{2 D t_0}} }{ c_0 s_0 } \left( \frac{t_B}{t_0} \right)^{5/2} = \begin{cases}
     \left( \frac{t_B}{t_0} \right)^{5/2} \mathscr{C} \left( \frac{t}{t_0}; \frac{s_0}{\sqrt{2Dt_0}}\right)
    &\quad \textrm{for} \quad t / t_B \leq \chi, \\
    \sqrt{5} q(\chi) e^{-\lambda_1 t / t_B} &\quad \textrm{for} \quad t / t_B > \chi.
    \end{cases}
    \label{eq:cmax_infplane_3}
\end{align}
Here we have defined the shorthand $q(\chi) = \chi^{-5/2} e^{\lambda_1 \chi}$.
Using the numerical data for the exponential decay in \cref{fig:infplate_panels}(\textit{c}), we may estimate $q$ directly.
In particular, reorganizing \cref{eq:cmax_infplane_3} for $t \gg t_B$, we estimate
\begin{align}
     q(\chi) \approx
    \frac{ {\sqrt{2 D t_0}} }{ c_0 s_0 \sqrt{5} } \left( \frac{t_B}{t_0} \right)^{5/2} \overline{ c_{\rm max}^{\rm num}(t)  e^{\lambda_1 t / t_B} },
\end{align}
where $c_{\rm max}^{\rm num}$ refers to the numerically obtained values and we use the arithmetic average $\overline{\bullet} = (t_2 - t_1)^{-1} \int_{t_1}^{t_2} \bullet \, \diff t$ in the interval $t \in [t_1, t_2]$.
Using $[t_1, t_2] = [2t_B, 5 t_B]$ yields $q \simeq 5.67$ and $\chi = 5/(2\lambda_1) W (2 \lambda_1 q^{-2/5} / 5 ) \simeq 0.41$, where $W$ is the Lambert W function. 

To find a unified scaling form for $c_{\rm max}$, we now introduce the scaled variable 
\begin{align}
    \xi  = \left( \frac{t_B}{t_0} \right)^{5/2} \mathscr{C} \left( \frac{t}{t_0}; \frac{s_0}{\sqrt{2Dt_0}}\right),
    \label{eq:xidef}
\end{align}
which, using \cref{eq:scrC_approx} becomes $\xi \simeq \sqrt{5} (t / t_B) ^{-5/2}$ for $t \gg t_0$.
The latter can be inverted to yield 
\begin{align}
    \frac{t}{t_B} \simeq \sqrt[5]{5} \xi^{-2/5} .
\end{align}
Together with \cref{eq:xidef}, this means that \cref{eq:cmax_infplane_3} can be written as
\begin{align}
    \frac{c_{\rm max} {\sqrt{2 D t_0}} }{ c_0 s_0 } \left( \frac{t_B}{t_0} \right)^{5/2} = \begin{cases}
     \xi & \quad \textrm{for} \quad \xi \geq \sqrt{5} \chi^{-5/2}, \\
    \sqrt{5} q(\chi) e^{-\lambda_1 \sqrt[5]{5} \xi^{-2/5} } &\quad \textrm{for} \quad \xi < \sqrt{5} \chi^{-5/2}.
    \end{cases}
    \label{eq:cmax_infplane_4}
\end{align}
Notably, the latter expression depends only on the single variable $\xi$.

As shown in \cref{fig:infplate_collapse}(\textit{b}), we find convincing data collapse by replotting our simulation data from \cref{fig:infplate_panels}(\textit{c}) as $c_{\rm max} \sqrt{2Dt_0} (t_B/t_0)^{5/2}/(c_0 s_0)$ against $\xi$ (\cref{eq:xidef}).
This suggests that a unified description for the maximum concentration is given by
\begin{align}
    c_{\rm max}(t) = \frac{c_0 s_0}{\sqrt{2 D t_0}}  \left( \frac{t_B}{t_0}\right)^{-5/2}
    H \left(
    \left( \frac{t_B}{t_0}\right)^{5/2}
    \mathscr{C} \left( \frac{t}{t_0} \right)
    \right)
    , \label{eq:cmax_unified_infplane}
\end{align}
where the scaling function $H(z) \to z$ for $z \gg 1$ and $H(z) \to \sqrt{5} q(\chi) e^{-\lambda_1 \sqrt[5]{5} z^{-2/5} } $ for $z \ll 1$, consistent with \cref{eq:cmax_infplane_4}.
In \cref{fig:infplate_collapse}(\textit{b}), we superimpose the function \eqref{eq:cmax_infplane_4} on top of the numerical data, showing that the latter describes the numerical data very well.

\subsection{Mixing regimes}
\label{sec:all_regimes_infplate}
\begin{figure}
    \centering
    \includegraphics[width=0.99\textwidth,trim={0.4cm 0.45cm 0.7cm 0.3cm},clip]{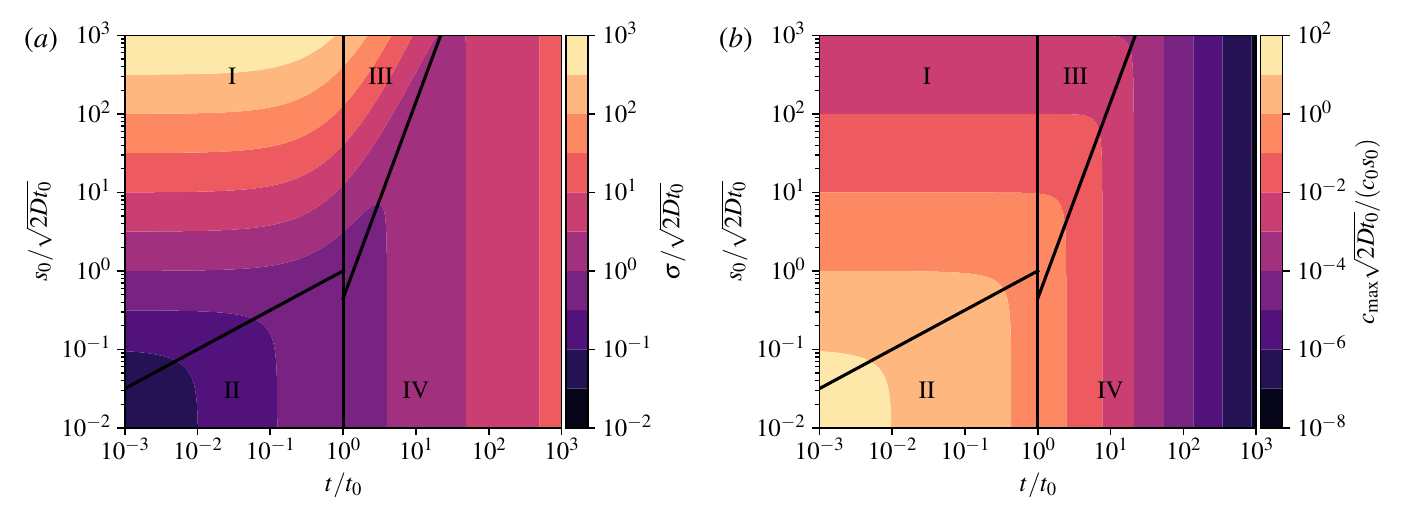}
    \caption{`Phase' diagram of mixing for a strip approaching an infinite plate before the saturation regime, showing (\textit{a}) the strip half-width $\sigma$ (\cref{eq:sigma_infplate}) and (\textit{b}) the maximum concentration $c_{\rm max}$ (\cref{eq:cmax1d}) as functions of the initial thickness $s_0$ and time $t$.
    Each numerical or experimental realization of a strip will follow horizontal lines from left to right in the diagram.
    The regimes I--IV are described in detail in \cref{sec:all_regimes_infplate} in the main text.
    }
    \label{fig:phase_diagram_infplane}
\end{figure}

We now have a quantitative description and a physical understanding of all relevant regimes for the case of an infinite plate.
There are four regimes (I-IV below) that can be obtained as different limits of the expressions for algebraic stretching, \cref{eq:sigma_infplate,eq:cmax1d}, and one regime (V) corresponding to exponential decay by wall interaction. 
The regimes can be summarized by:
\begin{subequations}
\begin{enumerate}[I.]
    \item Slow evolution of a thick strip (with $s_0 > 0$), $t \ll t_0, t_D = s_0^2 / (2D)$:
    \begin{align}
        \sigma(t) \simeq s_0 , 
        \quad
        c_{\rm max} (t) \simeq c_0.
    \end{align}
    \item Diffusive spreading, $t_D \ll t \ll t_0$:
    \begin{align}
        \sigma (t) \simeq \sqrt{2Dt}, \quad
        c_{\rm max}(t) \simeq \frac{c_0 s_0}{\sqrt{2Dt}}.
        \label{eq:sigmacmax_diffusivespreading_infplate}
    \end{align}
    \item Compression (with $s_0 > 0$), $t_0 \ll t \ll t_D = s_0^2 / (2D)$:
    \begin{align}
        \sigma(t) \simeq \frac{ s_0 }{\left[ 1 + \frac{t}{t_0} \right]^{2}}
        \stackrel{t \gg t_0}{\sim} t_0^2 s_0 t^{-2}, 
        \quad
        c_{\rm max} (t) \simeq c_0.
    \end{align}
    \item Asymptotic stretching, $t_0, t_D \ll t \ll t_B$:
    \begin{align}
        \sigma (t) \simeq \sqrt{\frac{2Dt}{5}}, \quad
        c_{\rm max}(t) \simeq \frac{c_0 s_0 \sqrt{5} t_0^{2} }{\sqrt{2 D} } t^{-5/2}.
        \label{eq:sigmacmax_asymptotic_stretching_infplate}
    \end{align}
    \item Asymptotic saturation and exponential decay, $t \gg t_B$:
    \begin{align}
        \sigma (t) \simeq 0.447 s_B , \quad
    c_{\rm max} (t) \simeq \frac{c_0 s_0 \sqrt{5} q(\chi)}{\sqrt{2 D t_0}} \left( \frac{t_B}{t_0}\right)^{-5/2} \exp\left( - \frac{\lambda_1 t}{t_B} \right).
    \end{align}
\end{enumerate}\label{eq:asymptotic_regimes}\end{subequations}
As we can see, the competition between the timescales $t_0$, $t_B$ and $t_D$ determines which scaling regimes are actually encountered. For the conditions considered here, typically, $t_0 < t_B$, while $t_D$ can vary by tuning $s_0$.
The regimes I-IV can be summarized in a `phase diagram' shown in \cref{fig:phase_diagram_infplane}, which essentially plots \cref{eq:sigma_infplate} in \cref{fig:phase_diagram_infplane}(\textit{a}) and \cref{eq:cmax1d} in \cref{fig:phase_diagram_infplane}(\textit{b}), with superimposed `phase boundaries' based on the limits summarized above.
A single simulation or experiment will follow horizontal cuts from left to right until a point is reached where $t \simeq t_B$, and the saturation behaviour occurs.

Although these exact results are valid for the highly idealized case of flow close to an infinite plate, we will see that the general behaviour carries over to more general velocity fields, as considered in the next section.

\section{Theory for general obstacles}
\label{sec:general}
We now generalize the theory from \cref{sec:infplane}, considering more general obstacle shapes and velocity fields.
Like before, we consider a strip (or sheet) approaching a stagnation point. 
We first present the generalized theory, based, as in \cref{sec:infplane}, on the lamellar theory \citep{Villermaux2019} whereupon we extend the theory to the saturation regime. 

To specify the problem, we consider a symmetric domain with a single obstacle, and flow along the separatrix (see \cref{fig:sphere}(\textit{a,b}) for the particular case of a sphere).
With this assumption, the separatrix forms a straight line, which is taken without loss of generality to be given by $y=z=0$. 
The stagnation point is taken to be at $\v x = (x, y, z) = (0, 0, 0)$, and $u = \v u \bcdot \hat{\v x} > 0$ so that the region of interest is in the domain $x < 0$.
We take the initial position of the strip/sheet to be at $x=x_0 < 0$.
Our key assumption is that the strip/sheet is oriented normally to the separatrix such that the main direction of compression 
is along the separatrix.
For more general, e.g., asymmetric, velocity fields or initial states, these assumptions should nevertheless hold true sufficiently close to a stagnation point.

\subsection{Lamellar theory}
In the framework of lamellar theory, briefly recalled in \cref{sec:app:lamella}, the concentration distribution along the separatrix is assumed to approach a Gaussian, \cref{eq:gaussianc_1}.
As in \cref{sec:infplane}, the challenge is to find the compression rate and secondly the warped time as a function of time, $\tau(t)$ (\cref{eq:taudef}), which give the half-width $\sigma$ (\cref{eq:sigmadef}) and the maximum concentration $c_{\rm max} (t)$ (\cref{eq:cmaxasft}).
These quantities together quantifies mixing at stagnation points.

The mean position $x$ of a strip/sheet segment advected along the separatrix is given by
\begin{equation}
    \d x t = u(x).
    \label{eq:dxdt}
\end{equation}
This allows us to express the position as a function of time, $x(t)$, which we can formally obtain
by integrating \cref{eq:dxdt} for a general space-dependent velocity field $u(x)$:
\begin{equation}
    t = \int_{x_0}^{x} \frac{\diff x'}{u(x')}.
    \label{eq:tasfx}
\end{equation}
Here we have used the initial condition $x(t=0) = x_0$.

The striation thickness $s$ of the line evolves according to the compression rate $\gamma$, which can be expressed in two ways:
\begin{equation}
    -\gamma = \frac{1}{s} \d{s}{t} = \pd{u}{x}.
    \label{eq:compress_0}
\end{equation}
We can write \cref{eq:compress_0} as
\begin{equation}
    \frac{1}{s} \d{s}{t} = \d{u}{x} (x(t)),
    \label{eq:compress_1}
\end{equation}
where the right-hand side is written as a total derivative because $u$ only depends (directly) on a single argument $x$.
Combining \cref{eq:dxdt,eq:compress_1}, we obtain
\begin{equation}
    \d{\ln s}{x} \d x t = \d{\ln s}{x} u = \d{u}{x} \quad \implies \quad \d{\ln s}{x} = \d{\ln u}{x},
\end{equation}
which can be solved to yield
\begin{align}
    \frac{s(t)}{s_0} = \frac{u(x)}{u(x_0)}.
    \label{eq:surel_0}
\end{align}
The simple relation \cref{eq:surel_0} represents a key step to proceed with the analysis, as it allows us to connect the temporal dependence of the striation thickness $s(t)$ (left hand side) to the spatial dependence of the velocity field $u(x)$ (right hand side).

Using \cref{eq:sigmadef,eq:surel_0}, the half-width can be expressed as
\begin{equation}
    \sigma(t) = \frac{u(x(t))}{u_0} \sqrt{ s_0^2 + 2 D \tau} .
    \label{eq:sigma_push}
\end{equation}
Herein, we can express the warped time using \cref{eq:taudef,eq:surel_0} as
\begin{align}
    \tau = u^2(x_0) \int_{0}^{t} \frac{\diff t'}{u^2(x(t'))} = u^2(x_0) \int_{x_0}^{x} \frac{\diff x'}{u^3(x')}.
    \label{eq:tauasfx}
\end{align}
where we have used \cref{eq:dxdt} in the last equality.
 
The challenge is then to solve and invert \cref{eq:tasfx} to obtain the position as a function of time, $x(t)$, to insert into $u(x)$ and as the upper integrand in \cref{eq:tauasfx} to get $\tau(t)$. Then the half-width $\sigma (t)$ is defined as a function of time through \cref{eq:sigma_push}, and the maximum concentration $c_{\rm max}(t) $ through \cref{eq:cmaxasft}. 

\subsubsection{Compact expressions}
As shorthand notation, it is useful to define the integral
\begin{align}
    F_n(x) &= \int_{x_{\rm ref}}^x u^{-n}(x') \, \diff x',
    \label{eq:Fndef}
\end{align}
with some arbitrary reference position $x_{\rm ref}$.
This allows us to express \cref{eq:tasfx,eq:tauasfx} as 
\begin{align}
    t &= F_1(x)-F_1(x_0) \quad \textrm{and} \\ 
    \tau &= u^2(x_0) \left[ F_3(x) - F_3(x_0) \right].
    \label{eq:taudef_1}
\end{align}
Since $u>0$ except \emph{at} stagnation points, $F_1$ is guaranteed to have an inverse, $F_1^{-1}$, on the relevant interval (although it is not generally available as a closed form expression). 
This allows us to express
\begin{align}
    x(t) = F_1^{-1}\left( F_1(x_0) + t \right)
    \label{eq:x_general}
\end{align}
and accordingly
\begin{align}
    \tau(t) = u^2(x_0) \left[ F_3(F_1^{-1}\left( F_1(x_0) + t \right)) - F_3(x_0) \right].
    \label{eq:tau_general}
\end{align}
Formally, this fully determines the evolution of the strip/sheet in the range of validity of the lamellar theory, i.e.\ where the concentration profile can be assumed be Gaussian across the normal direction. 
We note that the analysis holds for general steady velocity fields $u(x)$ along the separatrix.

The integrals and inverse functions involved in \cref{eq:tau_general} must in general be solved numerically.
However, in \cref{sec:theory_asympt} we elucidate certain limits where analytical progress can be made.

\subsection{Asymptotics}
\label{sec:theory_asympt}
General velocity fields $\v u = (u, v, w)$ can be Taylor expanded close to a no-slip surface. 
We choose the local frame $\v x = (x, y, z)$ with the origin at the surface and with $x$ as the wall-normal coordinate.
Since the no-slip condition requires $\v u = \v 0$ everywhere on the surface, the expansion of $\v u$ contains no term independent of $x$:
\begin{align}
    \v u = \v b(y, z) x + \mathcal{O}(x^2),
\end{align}
where $\v b = (b_x, b_y, b_z) = \partial \v u / \partial x |_{x=0}$ is the skin friction field \citep{lester2016chaotic}.
From the incompressibility condition $\div \v u = 0$, it follows that $\partial_x u = b_x + \mathcal O (x^2) = - x (\partial_y b_y + \partial_z b_z) + \mathcal O (x^2)$, such that the leading-order behaviour of the normal velocity is $u = - \tfrac{1}{2} x^2 (\partial_y b_y + \partial_z b_z) + \mathcal O(x^3) \sim x^2$.

Near no-slip stagnation points, velocity fields can thus generally be written in the form 
\begin{align}
    u(x) = \alpha x^2 + \mathcal{O}\left( (x/\ell)^3 \right),
    \label{eq:u_close_scaling}
\end{align}
where $|x| \ll \ell$, and $\ell$ is some characteristic length scale.
Like in \cref{sec:infplane}, $\alpha$ is a constant that depends on geometry and flow conditions.
Far from the stagnation point, $|x| \gg \ell$, we may write
\begin{align}
    u(x) \simeq u_\infty,
    \label{eq:u_close_u_infty}
\end{align}
where $u_\infty$ is the far-field velocity.
Typically, we can express 
\begin{align}
    \alpha = \kappa u_\infty/\ell^2,
    \label{eq:alpha_def}
\end{align}
where $\kappa = \mathcal O(1)$ is a shape-dependent numerical prefactor.

We assume that the velocity field can be expressed as 
\begin{align}
    u(x) = u_\infty g( -x/\ell),
    \label{eq:ux_g}
\end{align} 
where $g$ is some dimensionless function to be specified for each geometry.
We may then identify $\kappa$ by Taylor expanding $u(x)$ around $x=0$ and equating it with \cref{eq:u_close_scaling}:
\begin{equation}
	u (x) = \frac{1}{2} u''(0) x^2 + \mathcal O(x^3) = \frac{u_\infty}{2 \ell^2} g''(0) x^2 + \mathcal O(x^3) = \frac{u_\infty \kappa }{\ell^2} x^2 + \mathcal O(x^3).
        \label{eq:ux_general_exp}
\end{equation}
The last equality allows us to identify $\kappa = \frac{1}{2} g''(0)$.

\subsubsection{Small-$|x|$ or late-time limit}
We now seek the asymptotic solution when the filament comes asymptotically close to the stagnation point. This situation can equivalently mean $|x(t)| \ll \ell$ or $t \gg t_a$, where $t_a = \ell/u_\infty$ is the characteristic advective time scale based on the far-field velocity $u_\infty$.
Choosing $x_{\rm ref} = x_0$ such that $F_n(x_0)=0$, we may split the integral \cref{eq:Fndef} in two:
\begin{align}
    F_n(x) = \int_{x_0}^{\delta} u^{-n}(x') \diff x' + \alpha^{-n} \int_{\delta}^{x} (x')^{-2n} \diff x',
\end{align}
where we have chosen $\delta$ such that $|x| \ll \delta \ll \ell $.
Keeping only the dominant term, we obtain
\begin{align}
    F_n(x) = \alpha^{-n} \frac{-x^{-2n+1}}{2n-1},
\end{align} 
which for the expressions relevant to \cref{eq:tau_general} yields
\begin{align}
    F_1(x) &= \alpha^{-1} (-x)^{-1} , \\
    F_3(x) &= \frac{\alpha^{-3}}{5} (-x)^{-5}, \\
    F_1^{-1}(t) &= - \alpha^{-1}  t^{-1}.
\end{align}
Inserting these expressions into \cref{eq:tau_general,eq:x_general} yields, respectively,
\begin{align}
    \tau(t) &= \frac{u^2(x_0) \alpha^{2}}{5} t^{5},
    \label{eq:tau_asympt}
    \quad \text{and} \\
    x(t) &= - \alpha^{-1} t^{-1}.
    \label{eq:x_asympt}
\end{align}
For the half-width, we find by inserting \cref{eq:tau_asympt,eq:x_asympt,eq:u_close_scaling} into \cref{eq:sigma_push} (neglecting the term involving $s_0$ since $\tau$ will be arbitrarily large):
\begin{align}
    \sigma(t) = \sqrt{\frac{2 D t}{5}} ,
    \label{sec:sigma_gen_asympt}
\end{align}
which is identical to the asymptotic expression \eqref{eq:sigmacmax_asymptotic_stretching_infplate} for the infinite plate found in \cref{sec:infplane}. 
This demonstrates that diffusive spreading by a reduced diffusion coefficient is universal to no-slip stagnation points.
It appears as a direct consequence of the quadratic scaling \cref{eq:u_close_scaling} close to the stagnation point.

The maximum concentration is given by \cref{eq:cmaxasft} as
\begin{align}
    c_{\rm max}(t) = \frac{c_0 s_0 \sqrt{5}}{u_0 \alpha \sqrt{ 2 D }} t^{-5/2}.
\end{align}
When $|x_0| \ll \ell$ and subsequently $u_0 = \alpha x_0^2$, this is consistent with the asymptotic result \eqref{eq:sigmacmax_asymptotic_stretching_infplate} for the infinite plate in \cref{sec:infplane}.


Assuming, on the contrary, $|x_0| \gg \ell$, and inserting \eqref{eq:alpha_def} for $\alpha$, we obtain:
\begin{align}
    c_{\rm max}(t) = \frac{c_0 s_0 \sqrt{5} }{\kappa \ell  \sqrt{ 2 }} \Peclet^{1/2} \left( \frac{u_\infty t}{\ell} \right)^{-5/2}.
\end{align}
Here, we have introduced the Péclet number $\Peclet = u_\infty \ell / D$ that measures the strength of advection (based on the far-field velocity $u_\infty$) relative to diffusion.
The quantity in parentheses is the time $t$ normalized by the characteristic advective time $t_a$.

\subsubsection{Early times}
For the small-time asymptote, \cref{eq:Fndef} becomes, choosing $x_{\rm ref} = x_0$:
\begin{align}
    F_n (x) \simeq \frac{x - x_0}{u_0^n} .
\end{align}
Inserting this expression into \cref{eq:taudef_1} yields $\tau = t$, which in combination with \cref{eq:sigma_push,eq:cmaxasft} (neglecting the $s_0$ term for simplicity) yields regular diffusive behaviour, as expected:
\begin{align}
    \sigma(t) = \sqrt{2 D t} \quad \textrm{and} \quad
    c_{\rm max} (t) = \frac{c_0 s_0}{\sqrt{2 D t} }.
\end{align}
This result is also consistent with the expressions \eqref{eq:sigmacmax_diffusivespreading_infplate} for the infinite plate.

\subsection{Saturation, breakdown of lamellar theory and mixing scales}
The lamellar theory eventually breaks down when the Gaussian concentration profile of the lamella reaches a solid boundary, as discussed in \cref{sec:infplane_sat}.
In particular, this occurs when $\sigma (t_*) \sim |x(t_*)| \sim s_B$, i.e.\
using \cref{eq:sigma_push},
\begin{gather}
    \frac{u^2(x_*)}{u_0^2}  \left( s_0^2 + 2 D \tau(x_*) )\right) \sim x_*^2
\end{gather}
Realizing that $x_*$ will typically, at least for sufficiently high $\Peclet$, be close to the no-slip stagnation point, i.e.\ $|x_*| \ll \ell$, we may then use the asymptotic result derived in the previous section. 
We thus expect \cref{eq:x_sat,eq:sigma_sat} to hold for sufficiently high $\Peclet$.
Inserting \cref{eq:alpha_def} for $\alpha$ in the expressions for \cref{eq:stB_def} yields
\begin{subequations}
\begin{align}
    s_B = \ell \kappa^{-1/3} \Peclet^{-1/3} \\
    t_B = 
    t_a \kappa^{-2/3} \Peclet^{1/3}.
\end{align}\label{eq:stB_gen}\end{subequations}
Remarkably, we thus find a Batchelor scale $s_B \sim \Peclet^{-1/3}$ and a mixing time $t_B \sim \Peclet^{1/3}$ that coincide with the scalings for the regular shear flow \citep{souzy2018mixing,Villermaux2019}.
However, some important elements should be noted.
The definition of the Péclet number $\Peclet$ in shear flows used, e.g., by \cite{souzy2018mixing}, is based on the initial half-width $s_0$, whereas in our case it is only dependent on (in fact, proportional to) the characteristic length scale $\ell$ of the solid obstacle. 
Secondly, the mixing scales in shear flow are typically based on the \emph{first} point where diffusion and compression balance, analogously to how we defined $(t_m, s_m)$ in \cref{eq:tstar_infplate,eq:sstar_infplate}.
As such, the asymptotic decay of maximum concentration in shear flows occurs in the bulk and is thus algebraic, $c_{\rm max} (t) \sim t^{-3/2}$ \citep{Villermaux2019}, whereas in our case it is exponential due to the boundary interaction.
It should also be noted that the observed scaling is distinct from that related to stagnation points in the bulk (i.e., saddle points) where $t_B \sim \ln \Peclet$ and $s_B \sim \Peclet^{-1/2}$ \citep{Villermaux2019}.

For moderate $\Peclet$, $\mathcal O(x^3)$ corrections to the quadratic scaling of the velocity field for small $x$ (see \cref{eq:ux_general_exp}) modify the asymptotic values $\mean{x}_\infty$, $\sigma_\infty$ and $\lambda$.
In \cref{sec:app:finiteD_pert} we use a perturbation method to derive the approximate scalings
\begin{subequations}
\begin{align}
    \frac{\sigma_\infty}{s_B} &= 0.620 - 0.27 \omega \kappa^{-4/3} \Peclet^{-1/3} + O \left(\Peclet^{-2/3} \right), \\
    \frac{\mean{x}_\infty}{s_B} &= 0.447 - 0.34 \omega \kappa^{-4/3} \Peclet^{-1/3} + O \left(\Peclet^{-2/3} \right), \\
    \lambda &= 1.241 + 1.07 \omega \kappa^{-4/3} \Peclet^{-1/3} + O \left(\Peclet^{-2/3} \right).
\end{align}\label{eq:asymptotic_scales_1}\end{subequations}
Here, we have used the coefficient $\omega = g'''(0)/6$, where we recall that $g$ is a dimensionless function describing the velocity field, as implicitly defined in \cref{eq:ux_g}, and $\kappa = g''(0)/2$.
The values of these coefficients will generally depend on the flow configuration, but the form \eqref{eq:asymptotic_scales_1} is universal.
We note that this result is only expected to hold for sufficiently high $\Peclet$, since neglecting diffusion along the strip/sheet in \cref{eq:pdtc_finiteD} will be a poor approximation for lower $\Peclet$.
In particular, the correction term is smaller in magnitude than the leading term for $\Peclet \gtrsim |\omega|^3 / \kappa^4$, which for the spherical geometry studied in the next section (see \cref{eq:kappadef_sphere}) yields $\Peclet \gtrsim 10$.

Finally, assuming that the first point where diffusion and compression balance occurs sufficiently close to the solid boundary that $u \sim x^2$, it is possible  to infer the $\Peclet$-scaling of $(t_m, s_m)$.
Combining \cref{eq:sigma_push,eq:u_close_scaling,eq:tau_asympt,eq:x_asympt}, we obtain
\begin{align}
    \sigma^2 = \frac{s_0^2}{u_0^2 \alpha^{2} t^{4} }  + \frac{2 D t}{5},
\end{align}
which has a minimum at
\begin{subequations}
\begin{align}
    t_m &= \left( \frac{10 s_0^2}{u_0^2 \alpha^{2} D } \right)^{1/5} 
    = t_a 10^{1/5} \left( \frac{s_0  u_\infty }{u_0 \kappa \ell } \right)^{2/5} \Peclet^{1/5} , \\
    \sigma_m &= \frac{\sqrt{5} D^{2/5}}{10^{2/5}} \left( \frac{s_0}{u_0 \alpha } \right)^{1/5} 
    = \ell \frac{\sqrt{5} }{10^{2/5}} \left( \frac{s_0 u_\infty}{\kappa u_0  \ell} \right)^{1/5} \Peclet^{-2/5}.
\end{align}\label{eq:mixing_scales_alt}\end{subequations}
These scalings are consistent with \cref{eq:tstar_infplate,eq:sstar_infplate} for the infinite plate and distinct from those found for shear flows \citep{souzy2018mixing}.
We note that if the minimum occurs further from the obstacle, the functional form of $t_m, s_m$ would be more complicated.


\section{Application: Mixing near a spherical solid obstacle}
\label{sec:obstacles_typical}
So far we have made few assumptions on the velocity field, apart from the fact that the solute strip/sheet is always oriented normally to a straight separation line, which corresponds to a symmetric flow setup.
We now specify the velocity field and develop the theory by using known expressions for a canonical example of obstacles from the porous media literature, namely, an isolated sphere in open flow.

\subsection{Theory for flow over a sphere}
\label{sec:theory_sphere}
The setup we consider is now identical to that shown in \cref{fig:sphere}(\textit{a,b}).
The Stokes equations,
\begin{equation}
    \mu \laplacian \v u = \grad p,
    \quad \div \v u = 0,
    \label{eq:stokes}
\end{equation}
are straightforward to solve analytically for flow around a sphere (see, for example \cite[ch. 6.2]{langlois2014slow}).
Taking the mean flow to be oriented along the $x$ axis, the velocity field along the separatrix (see \cref{sec:app:sphere}) is given by
\begin{equation}
    u = u_\infty g\left( -\frac{x}{R} \right),
\end{equation}
where $R$ is the radius of the sphere, and the dimensionless function $g$ is defined by 
\begin{align}
    g(z) = \left( 1 + z \right)^{-3} z^2 \left( \frac{3}{2} + z \right).
    \label{eq:gzdef}
\end{align}

\subsubsection{Lamellar theory}
We now seek the evolution of the sheet in the lamellar regime, which is the case when the concentration distribution is well approximated by a Gaussian and does not interact significantly with the boundary.
The analytical expression \eqref{eq:gzdef} allows us to compute the integrals in \cref{eq:Fndef} exactly:
\begin{align}
    F_n(x) &= -\frac{ R}{ u_\infty^{n}} \int_{-x_0/R}^{-x/R} g^{-n}(z) \, \diff z 
    = \frac{R}{u_\infty^n } \left[ f_n\left( - \frac{x}{R} \right) - f_n\left( - \frac{x_0}{R} \right) \right], 
\end{align}
where the dimensionless functions $f_n$, $n=1,3$ are given by
\begin{align}
    f_1(z) &= \frac{2}{3 z} - z + \frac{1}{18} \log(3 + 2 z) - \frac{14}{9} \log(z), \\
\begin{split}
    f_3(z) &= \frac{1}{14580} \Bigg( \frac{864}{z^5} + \frac{7560}{z^4} + \frac{29760}{z^3} + \frac{71360}{z^2} + \frac{130880}{z} + 5\frac{ 41 + 28 z}{(3 + 2 z)^2} \Bigg) \\
   & \quad - z  - \frac{61}{4374} \log( 3 + 2 z ) - \frac{9872}{2187} \log z.
   \end{split}
\end{align}
We confirm that $f_1(z)$ is monotonically increasing and therefore has an inverse function $f_1^{-1}$ for $z > 0$. 
Since the equation is transcendental, this must in practice be found by a simple numerical procedure.
It follows that $F_1$ also has an inverse.

We can now express the position $x$ and the warped time $\tau$ formally by, respectively,
\begin{align}
x(t) = F_1^{-1}(t) &= - R f_1^{-1} \left( f_1\left(-\frac{x_0}{R}\right) + \frac{u_\infty t}{R} \right)
\label{eq:xt_sphere}
\end{align}
and
\begin{align}
    \tau(t) = \frac{R}{u_\infty} g^2\left( -\frac{x_0}{R} \right) \left[ f_3\left( f_1^{-1} \left( f_1\left(-\frac{x_0}{R}\right) + \frac{u_\infty t}{R} \right) \right) 
    - f_3\left( -\frac{x_0}{R} \right)
    \right].
\end{align}
This yields, using \cref{eq:sigma_push},
\begin{align}
    \sigma(t) &= \sqrt{2 D t_a} \mathscr{S}\left( \frac{u_\infty t}{R}; -\frac{x_0}{R}, \frac{s_0}{\sqrt{2 D t_a}} \right) ,
    \label{eq:Delta_push_2}
\end{align}
where we have defined the scaling function
\begin{align}
\mathscr{S}(\tilde t; \tilde x_0, \tilde s_0) &= g \left( f_1^{-1} \left( f_1\left( \tilde x_0 \right) + \tilde t \right) \right) \sqrt{ \frac{\tilde s_0^2}{g^2\left( \tilde x_0 \right)} + f_3\left( f_1^{-1} \left( f_1\left( \tilde x_0 \right) + \tilde t \right) \right) - f_3\left( \tilde x_0 \right) }.
\end{align}
For the maximum concentration, \cref{eq:cmaxasft}, we obtain similarly
\begin{align}
    c_{\rm max}(t) 
    &= \frac{c_0 s_0}{\sqrt{2 D t_a}} \mathscr{C} \left( \frac{t}{t_a} ; - \frac{x_0}{R} , \frac{s_0}{\sqrt{2Dt_a}}\right) ,
    \label{eq:cmax_sphere}
\end{align}
where we have defined the scaling function
\begin{align}
    \mathscr{C}(\tilde t; \tilde x_0, \tilde s_0 ) = \frac{1}{\sqrt{\tilde s_0^2 + g^2\left( \tilde x_0 \right) \left[ f_3\left( f_1^{-1} \left( f_1\left(\tilde{x_0}\right) + \tilde t \right) \right) 
    - f_3\left( \tilde x_0 \right)
    \right]}} .
\end{align}
For large times $t \gg t_a$, this expression simplifies to
\begin{align}
    \mathscr{C}\left( \frac{t}{t_a}; \frac{x_0}{R}, \frac{s_0}{\sqrt{2 D t_a}} \right) 
    &\simeq \frac{\sqrt{5}}{\kappa g\left( - \frac{x_0}{R} \right)} \left( \frac{t}{t_a} \right)^{-5/2}.
    \label{eq:C_sphere_approx}
\end{align}
This calculation completes our quasi-analytical computation of the lamellar theory near a circular obstacle.

\begin{figure}
    \centering
    \includegraphics[width=0.99\textwidth,trim={0.4cm 0.45cm 0.4cm 0.3cm},clip]{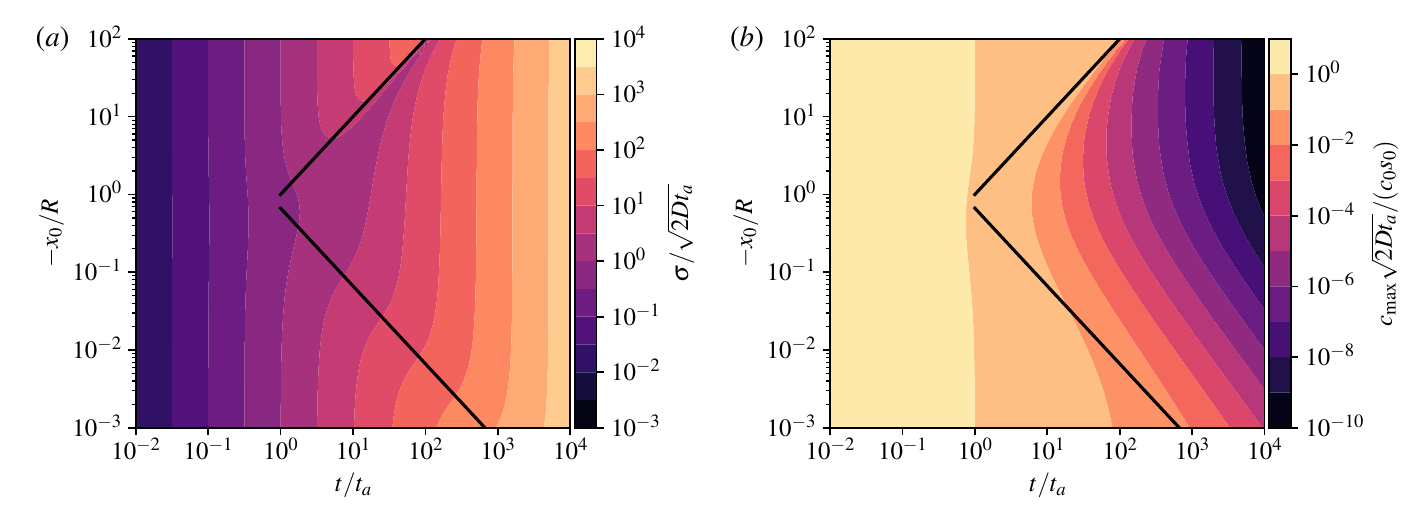}
    \caption{`Phase diagram' for the mixing of a solute sheet with initial thickness $s_0 \to 0$ approaching a stagnation point on a spherical obstacle. 
    We plot
    (\textit{a}) the strip half-width $\sigma$ (\cref{eq:Delta_push_2}) and (\textit{b}) the maximum concentration $c_{\rm max}$ (\cref{eq:cmax_sphere}) as functions of initial position $x_0$ and time $t$.
    Any experimental or simulated sheet of solute will evolve along horizontal lines in the diagram.
    The solid black lines indicate the transition from initial diffusive spreading to strong compression, $|x_0| \sim u_\infty t$ (for $|x_0| > R$), and from initial diffusive spreading to asymptotic stretching, $|x_0| \sim R^2 / (u_\infty t)$ (for $|x_0| < R$). }
    \label{fig:phase_diagram_sphere}
\end{figure}

The quasi-analytical theory derived above can be summarized in the diagram shown in \cref{fig:phase_diagram_sphere}, where we plot the half-width $\sigma$ (\cref{fig:phase_diagram_sphere}(\textit{a})) and the maximum concentration $c_{\rm max}$ (\cref{fig:phase_diagram_sphere}(\textit{b})) as functions of time $t$ and initial positions $x_0$ for infinitesimal initial thickness $s_0 \to 0$.
In particular, any sheet will evolve in time, i.e.\ along horizontal lines (constant $x_0$) in the diagram, until they reach the saturation stage specified in \cref{sec:saturation_sphere}.
As we can see from \cref{fig:phase_diagram_sphere}(\textit{a}), large initial distances $|x_0| > R$ lead to a long time in the (diffusive) widening regime, a subsequent strong compression regime, and finally a widening regime.
The corresponding maximum concentration drops sharply simultaneously with strong compression, as shown in \cref{fig:phase_diagram_sphere}(\textit{b}).
The large compression occurs when $|x_0| \sim u_\infty t$, i.e.\ when the strip has time to broaden significantly in the virtually compression-free region where $x > R$ and $u \simeq u_\infty$ before it reaches the strong compression region near the sphere.
This transition is shown as a solid line in \cref{fig:phase_diagram_sphere}.
For small separations, the velocity field is essentially quadratic, which leads to the same scaling as for the case of an infinite plate, as shown in \cref{fig:phase_diagram_infplane} in the $s_0 \to 0$ limit.
The transition between the two growth regimes can be found by using the results from \cref{sec:all_regimes_infplate}. 
In particular, the transition between regimes II and IV can be found when $t \sim t_0 = 1/(\alpha |x_0|) \sim R^2/(u_\infty |x_0|)$, which is also shown in \cref{fig:phase_diagram_sphere}.

\subsubsection{Saturation and asymptotic expressions}
\label{sec:saturation_sphere}
As mentioned previously, the lamellar theory breaks down when the solute sheet starts interacting with the solid obstacle, and we now investigate this regime for the case of a spherical obstacle.
Identifying the typical length scale $\ell=R$, we find from \cref{eq:gzdef}:
\begin{equation}
    \kappa = \frac{1}{2} g''(0) = \frac{3}{2} ,
    \quad
    \omega = \frac{1}{6} g'''(0) = -\frac{7}{2} .
    \label{eq:kappadef_sphere}
\end{equation}
The asymptotic results from the general model in \cref{sec:general} then carry over.
In particular, \cref{eq:kappadef_sphere} allows us to determine exactly the prefactors in \cref{eq:stB_gen} for the spherical geometry: 
\begin{subequations}
	\begin{align}
		s_B = R \left(\frac{3}{2} \right)^{-1/3} \Peclet^{-1/3}, \\
		t_B = 
		\frac{R}{u_\infty} \left(\frac{3}{2} \right)^{-2/3} \Peclet^{1/3},
  \label{eq:tB_sphere}
\end{align}\label{eq:stB_sphere}\end{subequations}
where we identify the Péclet number $\Peclet = u_\infty R / D$.
Furthermore, first-order corrections to the asymptotic scales $\mean{x}_\infty$, $\sigma_\infty$ and $\lambda$ are given by combining \cref{eq:asymptotic_scales_1,eq:kappadef_sphere}:
\begin{subequations}
\begin{align}
    \frac{\mean{x}_\infty}{s_B} &= 0.620 + 0.69 \Peclet^{-1/3} + O \left(\Peclet^{-2/3} \right), \label{eq:asymptotic_xmean_sphere} \\
    \frac{\sigma_\infty}{s_B} &= 0.447 + 0.54 \Peclet^{-1/3} + O \left(\Peclet^{-2/3} \right), \label{eq:asymptotic_sigma_sphere}\\
    \lambda &= 1.241 - 2.19 \Peclet^{-1/3} + O \left(\Peclet^{-2/3} \right).\label{eq:asymptotic_lambda_sphere}
\end{align}\label{eq:asymptotic_scales_1_sphere}\end{subequations}
In \cref{sec:simulations_sphere} below, specifically in \cref{fig:sigmainf_vs_Pe}, we compare \cref{eq:asymptotic_scales_1_sphere} both with numerical solutions using the exact velocity function $g$ and with values measured by particle tracking, finding good agreement.

\subsection{Validation by numerical simulations}
\label{sec:simulations_sphere}
We now test the theoretical predictions developed in \cref{sec:theory_sphere} against numerical simulations.
To do so, we use the particle simulation method described in \cref{sec:app:particle_simulation} with an initially infinitesimally thin strip ($s_0 \to 0$) starting at $x_0 = -R$ and run simulations with $N=10^6$ tracers at different Péclet numbers $\Peclet \in [10^3, 10^7]$.

\begin{figure}
    \centering
    \includegraphics[width=0.99\textwidth]{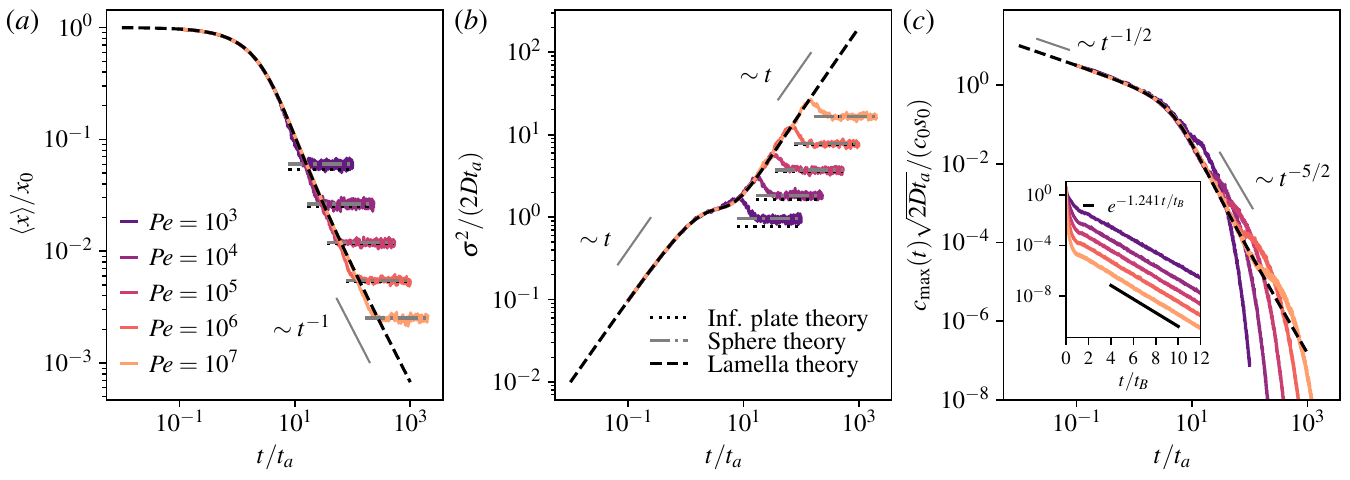}
    \caption{Evolution in time $t$ of solute sheets approaching a solid sphere for different Péclet numbers $\Peclet$.
    (\textit{a}) Mean position $\mean{x}$ of the strip, (\textit{b}) positional variance $\sigma^2$, and (\textit{c}) maximum concentration $c_{\rm max}$. The inset shows the same data in the main panel in a log-linear scale, revealing an exponential decay.
    The dashed line corresponds to expressions obtained using lamellar theory. 
    The simulated data for different $\Peclet$ follow the same curves which are predicted by our theory (\textit{a}) \cref{eq:xt_sphere}, (\textit{b}) \cref{eq:Delta_push_2} and (\textit{c}) \cref{eq:cmax_sphere}, and indicated by dashed black lines. This holds up to a $\Peclet$-dependent point where they saturate with a behaviour captured by first-order perturbation theory \cref{eq:asymptotic_scales_1_sphere}. 
    The lowest-order contribution, which is identical to the infinite-plate case (\cref{sec:infplane}) is less precise, but is increasingly accurate for the higher $\Peclet$ values.
    }
    \label{fig:sphere_panels}
\end{figure}
In \cref{fig:sphere_panels}(\textit{a}--\textit{c}) we plot, respectively, the mean position $\mean{x}$, the half-width $\sigma$ and the maximum concentration $c_{\rm max}$ as functions of time for the different simulations.
We compare these data to the theoretical predictions obtained using the lamellar theory in \cref{sec:theory_sphere} and to the asymptotic scales summarized in \cref{eq:asymptotic_scales_1_sphere}.
We see from \cref{fig:sphere_panels}(\textit{a}) that \cref{eq:xt_sphere} describes the simulated mean position $\mean{x}$ well until the saturation point, where the asymptotic value is captured by \cref{eq:asymptotic_xmean_sphere}. 
In contrast, the lowest-order approximation, which corresponds to the $\Peclet \to \infty$ limit or the case of an infinite plate (see \cref{eq:asymptotic_values_infplate}) slightly underpredicts the asymptotic value for lower $\Peclet$.
Similarly, \cref{fig:sphere_panels}(\textit{b}) shows that \cref{eq:Delta_push_2} well describes the simulated evolution of the half-width $\sigma$ until the saturation point and the subsequent plateau is described by \cref{eq:asymptotic_sigma_sphere}.
The same trend is observed here, that is, the infinite-plate approximation underpredicts the asymptotic value of $\sigma$, reflecting the fact that assuming a quadratic velocity field for all distances $x$ will tend to overestimate the compression rate.
\Cref{fig:sphere_panels}(\textit{c}) shows that there is good agreement between the simulated maximum concentration $c_{\rm max}$ and the theoretical expression \eqref{eq:cmax_sphere}, and the inset of the same figure shows that the tail decays exponentially, with an exponent close to $\lambda t/t_B$.

\begin{figure}
    \centering
    \includegraphics[width=0.99\textwidth]{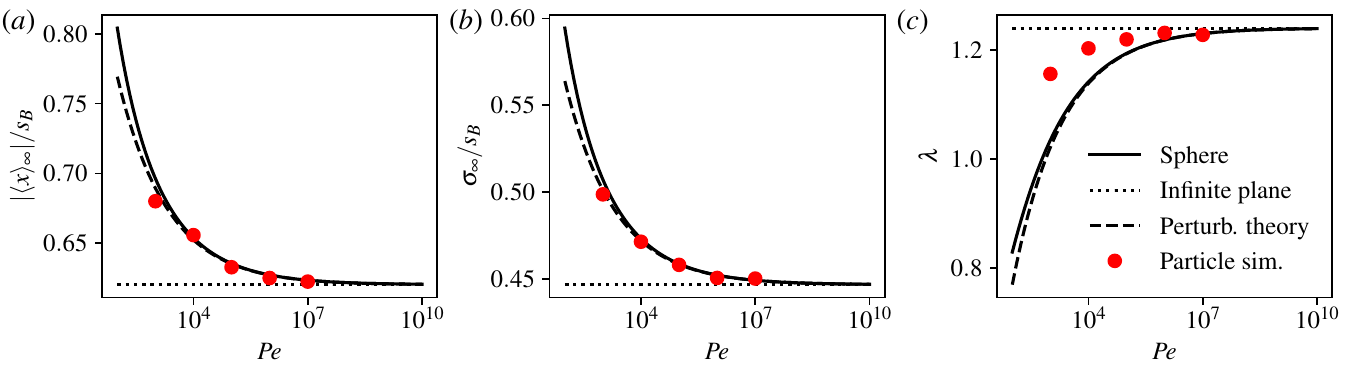}
    \caption{
    Comparison of key features of the asymptotic saturation state obtained by eigenfunction expansion using the full separatrix velocity field (solid line), the `infinite plate' velocity field, i.e.\ lowest-order expansion near the obstacle, (dotted line), first-order perturbation theory (dashed line), and particle simulations using the full velocity field.
    The method is detailed in \cref{sec:app:finiteD} and the analytical expressions are summarized in \cref{eq:asymptotic_scales_1_sphere}.
    The panels show the (\textit{a}) mean position $\mean{x}$, (\textit{b}) sheet half-width $\sigma$, and (\textit{c}) decay rate $\lambda$.
    As seen in (\textit{a}) and (\textit{b}) the agreement between the first-order theory and simulation data is excellent for the characteristic length scales.
    For the time scale shown in (\textit{c}), the agreement is satisfactory, but for lower $\Peclet$, diffusion along the sheet may start to play a role, accounting for some of the deviation.}
    \label{fig:sigmainf_vs_Pe}
\end{figure}

In \cref{fig:sigmainf_vs_Pe}, we test the numerically obtained saturation values against the values obtained by the eigenfunction expansion described in \cref{sec:app:finiteD}.
In particular, we compare the values for $\mean{x}_\infty$, $\sigma_\infty$, and $\lambda$ obtained by particle simulations (subject to the full 3D velocity field) to the corresponding values obtained (i) numerically using the full velocity field along the separatrix, (ii) by a first-order perturbation method (summarized in \cref{eq:asymptotic_scales_1_sphere} and described in \cref{sec:app:finiteD_pert}), and (iii) in the case of an infinite plate (described in \cref{sec:app:finiteD_infplate} and corresponding to the lowest-order contributions in \cref{eq:asymptotic_scales_1_sphere}).
Notably, we find in all cases very good agreement between the curves based on the eigenfunction expansion using the full velocity function and that using a perturbation method, revealing that the leading-order $\Peclet^{-1/3}$ corrections are in all cases sufficient in the current parameter range.
The particle simulation results align very well with these predictions for both $\mean{x}_\infty$ and $\sigma_\infty$ (shown in \cref{fig:sigmainf_vs_Pe}(\textit{a}) and (\textit{b}), respectively).
However, for $\lambda$, the agreement is somewhat less convincing and the simulated data appear to lie in the middle of the infinite plate and the two predictions based on more accurate velocity functions.
This can be attributed to the fact that the eigenfunction expansion method we have used assumes that transport is purely one-dimensional, while for low $\Peclet$, diffusion along the sheet will also contribute, leading to faster decay than what is predicted by the 1D model.
The agreement is nevertheless satisfactory and we use the data presented in \cref{fig:sigmainf_vs_Pe}(\textit{c}) as numerical evidence for using the simplifying approximation that $\lambda$ is a $\Peclet$-independent constant in the next section.

\subsection{Unified description and data collapse}
Having described the solute sheet evolution in all regimes, we now seek a unified description by general scaling functions.
For the strip width, we have from \cref{eq:Delta_push_2} that
\begin{align}
    \sigma = \begin{cases}
    \sqrt{2 D t_a} \mathscr{S} \left( \frac{t}{t_a}; -\frac{x_0}{R}, \frac{s_0}{\sqrt{2Dt_a}}\right)
    &\quad \textrm{for} \quad t \ll t_B, \\
    \sigma_\infty (\Peclet) &\quad \textrm{for} \quad t \gg t_B.
    \end{cases}
\end{align}
Dividing by $\sigma_\infty(\Peclet)$ and using \cref{eq:stB_sphere}, we obtain
\begin{align}
    \frac{\sigma}{\sigma_\infty(\Peclet)} = \begin{cases}
     \sqrt{2} \kappa^{1/3} \Peclet^{-1/6} \frac{s_B}{\sigma_\infty(\Peclet)} \mathscr{S} \left( \frac{t}{t_a}; -\frac{x_0}{R}, \frac{s_0}{\sqrt{2Dt_a}}\right)
    &\quad \textrm{for} \quad t \ll t_B, \\
    1 &\quad \textrm{for} \quad t \gg t_B.
    \end{cases}
    \label{eq:sigma_sphere_unified}
\end{align}
This expressions leads us to predict that the ratio $\sigma / \sigma_\infty(\Peclet)$ is a function of a single dimensionless variable:
\begin{align}
\psi = \kappa^{1/3} \Peclet^{-1/6} \frac{s_B} {\sigma_\infty(\Peclet)} \mathscr{S} \left( \frac{t}{t_a}; -\frac{x_0}{R}, \frac{s_0}{\sqrt{2Dt_a}}\right).
\label{eq:eta_def}
\end{align}
In the large-$\Peclet$ limit, we anticipate that $\sigma_\infty \to 0.447 s_B$, so that 
\begin{align}
    \sigma = s_B G \left( \kappa^{1/3} \Peclet^{-1/6} \mathscr{S} \left( \frac{t}{t_a}; -\frac{x_0}{R}, \frac{s_0}{\sqrt{2Dt_a}}\right) \right),
    \label{eq:sigma_G_sphere}
\end{align}
where $G$ is the same dimensionless function as defined in \cref{eq:unified_Sigma_infplate}.

\begin{figure}
    \centering
    \includegraphics[width=0.8\textwidth]{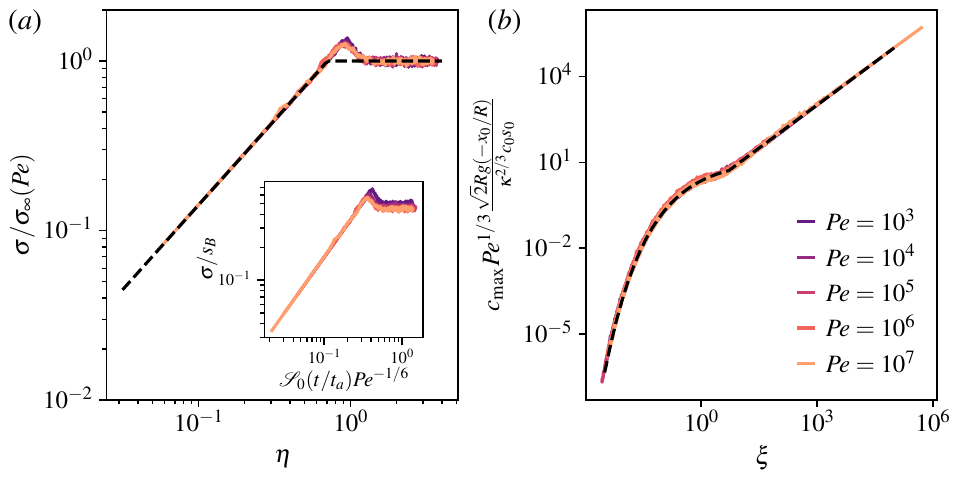}
    \caption{Data collapse of the same data as presented in \cref{fig:sphere_panels}.
    The colored solid lines represent simulations at different Péclet numbers $\Peclet$. 
    (\textit{a}) The simulated data for the half-width $\sigma$ normalized by $\sigma_\infty$ (\cref{eq:asymptotic_sigma_sphere}) collapses on a single curve when plotted as a function of the composite variable $\psi$ (\cref{eq:eta_def}) as predicted by \cref{eq:sigma_sphere_unified}. The master curve is close to the function $G(\psi) = \min(\sqrt{2}\psi, 1)$ which is superimposed as a dashed black line.
    The inset shows the same data plotted based on \cref{eq:sigma_G_sphere}, where $\mathscr{S}_0 (t/t_a) = \mathscr{S}(t/t_a; -x_0/R, s_0/\sqrt{2Dt_a})$, showing inferior agreement especially for lower $\Peclet$.
    (\textit{b}) The simulated data for the maximum concentration normalized by $s_0 c_0 \kappa^{2/3} \Peclet^{-1/3} /(\sqrt{2} R g(-x_0/R)$ plotted as a function of the composite variable $\xi$ (\cref{eq:xi_def_sphere}) falls on a single curve which can be described by the function $H(\xi)$ defined as the right hand side of \cref{eq:cmax_infplane_4}.
    }
    \label{fig:sphere_collapse}
\end{figure}

We now test the unified theoretical description for $\sigma$, \cref{eq:sigma_sphere_unified}, using the same simulation data as presented in \cref{fig:sphere_panels}(\textit{b}).
In particular, we plot $\sigma/\sigma_\infty$ as a function of the composite variable $\psi$ (\cref{eq:eta_def}), which \cref{eq:sigma_sphere_unified} predicts should fall on a single curve.
As seen in \cref{fig:sphere_collapse}(\textit{a}), the data collapse is excellent.
We further verify that the function $G (\psi) = \min(\sqrt{2}\psi, 1)$ captures the data in the low and high $\psi$ regimes.
In the inset of \cref{fig:sphere_collapse}(\textit{a}), we test the simplified expression \eqref{eq:sigma_G_sphere} by plotting the simulated $\sigma/s_B$ as a function of the dimensionless composite variable $\mathscr{S}(t/t_a; -x_0/R, s_0 / \sqrt{2Dt_a})\Peclet^{-1/6}$.
We find reasonable data collapse, except for the lower $\Peclet$ values, reflecting that for this range of $\Peclet$, the lowest order approximation of $\sigma_\infty$ is insufficient.

For the maximum concentration, we anticipate the following relation:
\begin{align}
    \frac{c_{\rm max} \sqrt{2 D t_a}}{s_0 c_0} = \begin{cases} 
        \mathscr{C}\left( \frac{t}{t_a}; - \frac{x_0}{R}, \frac{s_0}{\sqrt{2 D t_a}} \right) \quad \textrm{for} \quad t \ll t_B, \\
        K \exp\left( - \lambda_1 t / t_B \right) \quad \textrm{for} \quad t \gg t_B,
    \end{cases}\label{eq:cmax_cases_sphere}
\end{align}
where $K$ is an undetermined constant with respect to time, i.e.\ a function of $x_0/R$ and $s_0/\sqrt{2 D t_a}$.
For analytical convenience and because it was shown numerically to be a good approximation in \cref{sec:simulations_sphere}, we approximate $\lambda$ to be constant and equal to the value for an infinite plate, which corresponds to the $\Peclet \to \infty$ limit of \cref{eq:asymptotic_lambda_sphere}. 
To simplify \cref{eq:cmax_cases_sphere}, we extrapolate the two branches of the function and require them to be equal at a time $t_\crossproduct = \chi t_B \sim t_B$:
\begin{align}
    K \exp\left( - \lambda_1 \chi \right) = \mathscr{C}\left( \chi \frac{t_B}{t_a}; -\frac{x_0}{R}, \frac{s_0}{\sqrt{2 D t_a}} \right) 
    \simeq \frac{1}{g\left( - \frac{x_0}{R} \right)} \frac{ \sqrt{5}}{\kappa } \chi^{-5/2} \left( \frac{t_B}{t_a} \right)^{-5/2}.
\end{align}
where we have used \cref{eq:C_sphere_approx} and we recall from \cref{sec:unified_infplate} that $\chi \simeq 0.41$ and $q = \chi^{-5/2} e^{\lambda \chi } \simeq 5.67$.
Simplification yields
\begin{align}
    K = \frac{\sqrt{5} \kappa^{2/3} \Peclet^{-5/6} }{ g\left( - \frac{x_0}{R} \right)} q(\chi),
\end{align}
where we have used \cref{eq:tB_sphere}. Inserting for $K$ and reorganizing terms, we obtain
\begin{align}
    \frac{c_{\rm max} \sqrt{2} R }{s_0 c_0} g\left( - \frac{x_0}{R} \right) \kappa^{-2/3} \Peclet^{1/3}     
    = \begin{cases} 
        \kappa^{-2/3} \Peclet^{5/6} g\left( - \frac{x_0}{R} \right) \mathscr{C}\left( \frac{t}{t_a}; -\frac{x_0}{R}, \frac{s_0}{\sqrt{2 D t_a}} \right) \quad \textrm{for} \quad t \ll t_B, \\
        \sqrt{5} q(\chi) \exp\left( - \lambda_1 t / t_B \right) \quad \textrm{for} \quad t \gg t_B,
    \end{cases}
\end{align}
where we have used $\sqrt{D t_a} = R \Peclet^{-1/2}$.
We define the composite variable
\begin{align}
    \xi = \kappa^{-2/3} \Peclet^{5/6} g\left( - \frac{x_0}{R} \right) \mathscr{C}\left( \frac{t}{t_a}; -\frac{x_0}{R}, \frac{s_0}{\sqrt{2 D t_a}} \right),
    \label{eq:xi_def_sphere}
\end{align}
which for large $t$ becomes, again using \cref{eq:C_sphere_approx},
\begin{align}
    \xi \simeq \sqrt{5} \left( \frac{t}{t_B} \right)^{-5/2}
    \Leftrightarrow
    \frac{t}{t_B} \simeq 
    \sqrt[5]{5} \xi^{-2/5} .  
\end{align}
We thus find
\begin{align}
    \frac{c_{\rm max} \sqrt{2} R }{s_0 c_0} g\left( - \frac{x_0}{R} \right) \kappa^{-2/3} \Peclet^{1/3} = \begin{cases} 
        \xi &\quad \textrm{for} \quad \xi \gg \sqrt{5}, \\
        \sqrt{5} q(\chi) \exp\left( - \lambda_1 \sqrt[5]{5} \xi^{-2/5} \right) &\quad \textrm{for} \quad \xi \ll \sqrt{5}.
    \end{cases}
    \label{eq:cmax_almost_unified}
\end{align}
We realize that the quantity on the right hand side is identical to that in \cref{eq:cmax_infplane_4}.
Hence, we expect
\begin{align}
   c_{\rm max} 
    =  \frac{s_0 c_0 \kappa^{2/3} \Peclet^{-1/3} }{g\left( - \frac{x_0}{R} \right) \sqrt{2} R } 
    H \left( \kappa^{-2/3} \Peclet^{5/6} g\left( - \frac{x_0}{R} \right) \mathscr{C}\left( \frac{t}{t_a}; -\frac{x_0}{R}, \frac{s_0}{\sqrt{2 D t_a}} \right) \right),
    \label{eq:cmax_sphere_unified}
\end{align}
where the scaling function $H$ is the same as the one in \cref{eq:cmax_unified_infplane}.

In \cref{fig:sphere_collapse}(\textit{b}) we test the unified theoretical expression for $c_{\rm max}$, \cref{eq:cmax_sphere_unified}, by using the same data as presented in \cref{fig:sphere_panels}(\textit{c}).
We plot the simulated values of $c_{\rm max}$ multiplied by the factor ${g\left( - \frac{x_0}{R} \right) \sqrt{2} R } / (s_0 c_0 \kappa^{2/3}) \Peclet^{1/3}$ (that is, the quantity on the left-hand side of \cref{eq:cmax_almost_unified}) against the composite variable $\xi$ (\cref{eq:xi_def_sphere}) and obtain excellent data collapse.
In particular, the data for different $\Peclet$ fall on the same master curve, and this curve can be accurately described by the function $H(\xi)$ even in the transition region between the two regimes. 
This close agreement establishes confidence in \cref{eq:cmax_sphere_unified} and completes our unified description of mixing near stagnation points. 
We note that while these scaling forms are universal, they are not simple closed-form expressions and their practical utility lie in their limits, which are closed-form.

\section{Discussion and conclusion}
\label{sec:discussion}
In this work, we have developed a complete description of mixing of solute strips or sheets near no-slip stagnation points.
Combining the Lagrangian stretching framework and an eigenfunction expansion using general velocity fields evaluated along the separatrix, we have provided expressions for the maximum concentration and the half-width of these solute structures, two key quantities for quantifying scalar decay.
We have described asymptotic scaling regimes as these quantities evolve in time, summarized in \cref{eq:asymptotic_regimes}. 
We identified characteristic mixing time and length scales (\cref{eq:stB_gen,eq:mixing_scales_alt}), and uncovered how they scale with $\Peclet$.
In particular, we find the asymptotic length scale $s_B \sim \Peclet^{-1/3}$ and time scale $t_B \sim \Peclet^{1/3}$. While the same exponents can be found in other systems, as discussed both previously and below, the mechanisms by which they arise in this system have to the best of our knowledge not been described before.

The results presented here are relevant to a broad range of transport problems in which similar physics applies.
Laminar heat transfer of shear flow over a plate is a classical problem \citep{lienhard2020heat}.
The thermal boundary layer thickness $\delta_t$ that develops when the heat flux or temperature is maintained in the plate is expected to scale with the velocity boundary layer thickness $\delta_v$  as 
\begin{equation}
    \frac{\delta_t}{\delta_v} = \Prandtl^{-1/3}
\end{equation}
where the Prandtl number $\Prandtl=\nu/D_t$ is defined as the ratio of the kinematic viscosity $\nu$ to the thermal diffusivity $D_t$.
This result is comparable to the solute boundary layer thickness obtained near a stagnation point, $s_B=(D/\alpha)^{1/3}$. 
In fact, taking the velocity length scale as $\delta_v \sim (\nu/\alpha)^{1/3}$, one recovers the previous scaling, i.e., $\delta_t \sim (D_t / \alpha)^{1/3}$.
It is interesting to note that two different flows, shear flow and stagnation point flow, both lead to the same scalar transport problem close to the boundary. 
This behaviour is likely due to the fact that shear and compression are intimately connected through the incompressibility condition, $\div \v u = 0$, which, sufficiently close to a stagnation point, makes both shear and compression proportional to $\alpha$, and thus proportional to each other.
This is particularly clear for the flow field close to an infinite plate, cf. \cref{eq:u_infplate_3d}.
That said, the two mixing problems still differ through the existence of a fixed mixing scale in the case of the stagnation point, while shear cannot balance the effect of diffusion at long times. 
In other words, a pulse of heat injected close to a plate sheared by a cold fluid flow would see its characteristic length scale growing with time (as $\sqrt{D_t t}$), while the same pulse of heat impinging on a wall would be blocked at the scale $s_B$.


The problem of heat transfer close to a wall is analogous to the reactive transport problem between a fluid and a solid. 
An infinitely fast first-order reaction between a mixing solute A in the fluid and the solid is equivalent to a conservative transport problem for A with 0 concentration on the solid boundary. 
The thickness of the depletion layer in the fluid due to the absorption of A on the solid is thus expected to scale as $D^{1/3}$, regardless of the local boundary flow topology (shear or stagnation point flow). 
This scaling was observed by \citet{aquino2023fluid} in 3D porous media flows and by \citet{gouillart2007walls} in 2D chaotic mixers with walls.
\citet{aquino2023fluid} attributed the 1/3 exponent to the shear velocity profile. 
Here, we have demonstrated that the presence of stagnation points on the surface of beads can also produce a similar depletion layer.

By including solute interaction with the solid boundary in the description, we have provided a first step towards extending lamellar models to include boundary interaction, which is currently missing in these models \citep{meunier2010diffusive,martinez2018diffusive}.
Future work should therefore aim to extend the description of boundary interaction beyond stagnation points. 
The theory developed for stretching along streamlines in steady 2D velocity fields \citep{dentz2016coupled,pierce2026pore} and 3D Darcy flows \citep{lester2018fluid} would be natural starting points for extending concentration dynamics into 2D and 3D velocity fields.

The general theory was applied to two canonical examples of viscous flow: flow around a spherical obstacle and flow towards an infinite plate.
It is important to note that the general analysis presented is valid for any flow field evaluated along the separatrix, as long as the solute strip/sheet is oriented perpendicularly to this line. 
The expressions derived here allow for direct experimental verification across a wide range of flow conditions, provided that several orders of magnitude in local concentrations can be resolved \citep{deanna2014mixing}.

Although further investigations could consider laminar flow at higher Reynolds number, it is unlikely that the asymptotic behaviour would be affected, as the flow will be arbitrarily slow near the solid boundary.
However, at moderate Reynolds numbers, inertial recirculation could emerge and alter the mixing and reaction dynamics \citep{chen2024inertia}.
Fluids with more complex rheology could potentially modify the scaling \citep{taylor2025stagnation}.
Moreover, considering different physical conditions, such as boundary slip (which would reduce the asymptotic scaling to the same as in open saddle flows \citep{Villermaux2019}).
Natural avenues for extending the present research could be towards
multiphase flow \citep{mathiesen2023dynamic,velasquez2022sharp,linga2026dynamic},
time-dependent flow fields \citep{oxaal1994irreversible,flekkoy1995hydrodynamic,Heyman2021,fiori2025solute}, multiple interacting lamellae \citep{souzy2018mixing}, continuous solute injection scenarios \citep{sole2023closer,borgman2023solute,farhat2025linking,hallack20253d}, fluid-fluid reactions \citep{shafabakhsh2024resolving,yoon2021optimal}, and different boundary conditions relevant for fluid-solid reactions \cite{steefel2005reactive}.
In particular, stagnation points have shown relevance for reactions in inviscid flows \citep{conti1969inviscid} and for the control of chemical waves by fluid stretching and compression \citep{izumoto2024control}.
Recent theoretical progress on line stretching statistics \citep{pierce2026pore,lester2025line} provides tools for connecting local stretching at stagnation points to the global stretching distribution.
In porous media flow fields, stagnation points and the separatrices connecting them are ubiquitous, but they are also objects of zero measure.
The theory developed in this paper is valid in a finite neighborhood around these structures, and so an important challenge for future work is to investigate the consequences of mixing near stagnation points for transport at macroscopic scales.



\begin{bmhead}[Acknowledgements]
The authors thank E.~G.~Flekkøy and K.~Pierce for useful discussions.
Generative AI tools (Anthropic Claude) were used for basic language editing and formatting. All scientific content, derivations, results and references are the authors' own.
\end{bmhead}

\begin{bmhead}[Funding]
{This work was supported by the Research Council of Norway through its Centers of Excellence funding scheme (project no. 262644, PoreLab CoE) and its FRIPRO scheme (G.L., project no. 325819, M4; T.L.B., project no. 353372, MinMix).}    
\end{bmhead}

\begin{bmhead}[Declaration of interests]
{The authors report no conflict of interest.}
\end{bmhead}


\begin{bmhead}[Author ORCIDs]
{
G.\ Linga, https://orcid.org/0000-0002-0987-8704;
T.\ Le Borgne, https://orcid.org/0000-0001-9266-9139;
J.\ Heyman, https://orcid.org/0000-0002-0327-7924
}    
\end{bmhead}


\begin{appen}

\section{Brief summary of the lamellar theory}
\label{sec:app:lamella}
We here briefly recall the central steps of the Lagrangian stretching framework applied to a strip/sheet \citep{ranz1979applications,meunier2010diffusive,Villermaux2019}.
We consider a moving reference frame such that $x' = x - \bar x(t)$ points in the direction normal to the strip/sheet centered at $x = \bar x(t)$. 
Neglecting diffusion along the strip, and linearizing the velocity field, the advection--diffusion \cref{eq:ade3d} simplifies to 
\begin{align}
    \partial_t c = -\gamma x' \partial_{x'} c + D \partial_{x'}^2 c,
    \label{eq:pdtc_lin}
\end{align}
where $\gamma = - \dot s / s$ is the compression rate at $x = \bar x(t)$
and $s(t)$ is the {striation thickness} (equivalent width of the distribution in the absence of diffusion), where $s(0) = s_0$.
Introducing the dimensionless spatial coordinate $\zeta = x' / s(t)$ and the {warped time} $\tau$ through \cref{eq:taudef}, the governing \cref{eq:pdtc_lin} reduces to the simple diffusion equation $\partial_\tau c = D s_0^{-2} \partial_\zeta^2 c$.
For sharply peaked initial conditions, the latter admits the classical Gaussian solution
\begin{align}
    c(\zeta, \tau) = \frac{c_0}{\sqrt{1 + \frac{2 D}{s_0^2} \tau}} \exp \left( - \frac{\zeta^2}{1 + \frac{2 D} {s_0^2} \tau} \right),
    \label{eq:gaussianc_warped}
\end{align}
where the initial condition is given by $c(\zeta, 0) = c_0 \exp(-\zeta^2)$.
Transforming back to physical coordinates, the lamellar theory \citep{ranz1979applications,Villermaux2019} predicts the Gaussian concentration distribution around it as
\begin{align}
    c (x, t) = c_{\rm max}(t) \exp\left( - \frac{(x-\bar x(t) )^2}{2\sigma^2(t)} \right),
    \label{eq:gaussianc_0}
\end{align}
where we identify 
\begin{align}
    c_{\rm max}(t) = \frac{c_0}{\sqrt{1 + \frac{2 D}{s_0^2} \tau(t)}} \quad \text{and} \quad
    \sigma (t) = s(t) \sqrt{ 1 + \frac{2 D}{s_0^2} \tau(t)}.
\end{align}

\section{Particle simulations}
\label{sec:app:particle_simulation}

To solve the advection-diffusion equation \eqref{eq:ade3d}, we use a method based on tracking diffusive particles, implemented in the code \textsc{Partrac} \citep{linga2026partrac}. 
From the distribution of these particles, we can reconstruct the concentration field.
The advantage of this method is that it does not depend on a discretization of space and is free from numerical diffusion. 
Thus it can be used for arbitrarily thin strips/blobs at very high Péclet number.

For each particle $i$, we solve the Langevin equation,
\begin{equation}
    \d {\v x_i}{t} = \v u(\v x_i) + \gv \eta_i(t),
\end{equation}
to obtain the position $\v x_i$ as a function of time $t$, where $\v u(\v x)$ is the analytical velocity field, and $\gv \eta_i(t)$ is uncorrelated Gaussian random noise with the properties
\begin{align}
\mean{\gv \eta_i (t)} &= \v 0, \\
\mean{\gv \eta_{i} (t) \otimes \gv \eta_{j} (t') } &= 2 D \delta_{ij} \delta(t - t') \mathsfbi{I}_d.
\end{align}
Here, $D$ is the diffusion coefficient, $\delta_{ij}$ is the Kronecker delta, and $\delta(t)$ is the Dirac delta distribution.

The disadvantage of the method is that it is less precise than Eulerian methods for calculating concentration fields \citep{shafabakhsh2024resolving}.
In particular, a large number of particles is required to resolve local concentration variations.
The precision is particularly vulnerable in low concentration regions, which will eventually occur for scalar decay near a solid boundary.

To resolve this issue, we exploit the fact that we are mainly interested in the concentration in a part of the domain close to the separatrix, which we call the region of interest.
We assign a weight $w_i$ (i.e., mass) to every particle $i$.
When a particle exits the region of interest, which in this case should be downstream of the stagnation point, that particle is removed, and a random particle $i$ within the region of interest is `split' into two, such that the number of particles $N$ is conserved.
The weights of both the old particle $i$ and the new particle $j$ are then set to $w_i, w_j \leftarrow w_i/2$.
The concentration distribution and the spatial moments can be reconstructed using the positions $\v x_i(t)$ and the weights $w_i(t)$.
The method is implemented in the module \texttt{weighted\_walkers} in \textsc{Partrac} \citep{linga2026partrac}.

\subsection{Positional moments}
We estimate the positional moments ($q=1,2, \ldots$) along the separatrix by:
\begin{align}
    \mean{x^q} = \frac{\sum_{i} w_i x_i^q  \Theta (\v x_i)}{\sum_i w_i \Theta(\v x_i)},
    \label{eq:xq}
\end{align}
where $\Theta$ is an indicator function defined such that $\Theta(\v x) = 1$ when $\v x$ is in a thin region around the separatrix (defined below) where we compute moments, and $\Theta(\v x) = 0$ otherwise. 
We can then obtain the mean position of the strip/sheet $\mean{x}$ directly from \cref{eq:xq} and the half-width as 
\begin{align}
    \sigma = \sqrt{\mean{x^2}-\mean{x}^2}.
\end{align}
In particular, we use 
\begin{align}
    \Theta(\v x) = \begin{cases}
        H (\Delta y / 2 - |y|) &\quad \text{(sphere),} \\
        H (\Delta \varrho - \sqrt{y^2 + z^2})       &\quad \text{(infinite plate),}
    \end{cases}
\end{align}
where $H(\bullet)$ is the Heaviside step function.
Here, $\Delta y$ is the size of a small interval around $y=0$ which we use to probe averages along $x$; $\Delta \varrho$ is the radius of a small interval around $\varrho = \sqrt{y^2 + z^2} = 0$.

\subsection{Concentration fields for the infinite plate}
To obtain concentration fields from particle simulations, we reconstruct the local concentrations by spatially binning particle positions.
For the case of the infinite plate, where the flow is effectively two-dimensional, we sum over the set of particles $P(\Omega_j)$ inside the small box $\Omega_j$ defined by
\begin{align}
    \Omega_j := 
    (x, y) \in [x_j - \Delta x/2, x_j + \Delta x/2] \crossproduct [y_j - \Delta y/2, y_j + \Delta y/2]. 
\end{align}
Here, $\Delta x \crossproduct \Delta y$ are the $(x,y)$-dimensions of the box and we count all particles in the $z$ direction.
The concentration in the cell $\Omega_j$ is then
\begin{align}
    c(\v x) \approx C \sum_{i \in P(\Omega_j)} w_i = C W_j, \quad \v x \in \Omega_j,
\end{align}
where $w_i$ is the weight of particle $i$ and $C$ is a scaling factor related to the initial maximum concentration of the sheet/strip, $c_{0}$, and its initial thickness, $s_0$, to be specified below.
Integrating over $xy$-space, i.e.\ summing over all cells, we obtain the total mass per depth:
\begin{align}
    \mathscr M = \int c(\v x) \, \diff^2 \v x \approx 
    C \Delta x \Delta y \sum_j W_j = C \Delta x \Delta y N,
\end{align}
where $N$ is the total number of particles. 
To obtain  $C$, we can also express $\mathscr M$ directly from the initial condition \eqref{eq:c_0_x},
\begin{align}
    \mathscr M = \ell \int_{-\infty}^\infty c_0(x) \, \diff x = \ell c_{\text{max}, 0} \int_{-\infty}^\infty \exp \left( \frac{(x - x_0)^2}{2 s_0^2} \right) \, \diff x = \ell c_{\text{max},0} s_0 \sqrt{2\pi},
\end{align}
where $\ell$ is the initial length in the $y$ direction. The normalization factor is then obtained as
\begin{align}
    C = \frac{\ell c_{\text{max},0} s_0 \sqrt{2\pi}}{\Delta x \Delta y N}.
\end{align}
The scaled concentration then becomes
\begin{align}
    \frac{c(\v x) \sqrt{2 D t_a}}{c_{\text{max}, 0} s_0} = \frac{2 \ell \sqrt{\pi D t_a} }{\Delta x \Delta y N} W_j .
\end{align}

\subsection{Concentration fields for the sphere}
The fluid velocity field around a sphere is three-dimensional and cylindrically symmetric around the separatrix, so to obtain the concentration fields we average over the azimuthal coordinate $\theta$. We consider the subdomains
\begin{align}
    \Omega_j := \{ (x, \varrho, \theta) \in [x_j - \Delta x/2, x_j + \Delta x/2] \crossproduct [r_j - \Delta r/2, r_j + \Delta r/2] \crossproduct  [0, 2 \pi] \},
\end{align}
which allows to express the local concentrations
\begin{align}
    c (\v x) = \sum_{i \in P(\Omega_j)} C(\varrho_i) w_i,
    \quad \text{where} \quad
    C(\varrho) = \frac{C'}{\varrho_i},
\end{align}
and $\varrho_i$ is the distance from the $x$-axis to the volume element, i.e.\ $\varrho_i = \sqrt{y_i^2 + z_i^2}$.
Summing over all particles, this yields the total mass
\begin{align}
    \mathscr M \approx 2 \pi \Delta r \Delta x \sum_j \sum_{i \in P(\Omega_j)} \varrho_i C (\varrho_i) w_i
    = 2 \pi \Delta r \Delta x N C' 
    \approx \pi \ell^2 c_{\text{max},0} s_0 \sqrt{2\pi},
\end{align}
where the right hand side is obtained by integrating over the initial distribution, and $\ell$ is the radius of the initial circular sheet.
We obtain the normalization factor
\begin{align}
    C' = \frac{\ell^2 c_{\text{max},0} s_0 \sqrt{\pi/2}}{\Delta r \Delta x N}.
\end{align}
The scaled concentration becomes
\begin{align}
    \frac{c(\v x) \sqrt{2 D t_a}}{c_{\text{max}, 0} s_0} = \frac{\ell^2 \sqrt{\pi D t_a}}{\Delta r \Delta x N}  \sum_{i \in P(\Omega_j)} \frac{w_i}{\varrho_i} .
\end{align}

\section{Stokes flow around a sphere}
\label{sec:app:sphere}
The analytical solution to Stokes equations \eqref{eq:stokes} around a sphere is standard textbook knowledge, see e.g.\ \cite[ch. 6.2]{langlois2014slow}. 
For a sphere of radius $R$, the velocity field is given in spherical coordinates by
\begin{subequations}
\begin{align}
    u_r &= u_\infty \left( 1 - \frac{3}{2} \left( \frac{r}{R} \right)^{-1} + \frac{1}{2} \left( \frac{r}{R} \right)^{-3} \right) \cos \theta, \\
    u_\theta &= u_\infty \left( - 1 + \frac{3}{4} \left( \frac{r}{R} \right)^{-1} + \frac{1}{4} \left( \frac{r}{R} \right)^{-3} \right) \sin \theta,
\end{align}\label{eq:u_rtheta_sphere}\end{subequations}
where $\theta$ measures the polar angle with respect to the flow direction.
Taking the flow direction to be along $x$, along the separatrix ($y=z=0$), we easily find
\begin{align}
    u_x = u_\infty \left( 1 - \frac{3}{2} \left( \frac{r}{R} \right)^{-1} + \frac{1}{2} \left( \frac{r}{R} \right)^{-3} \right).
\end{align}
To express the field in terms of the coordinate $x$ along the separatrix, with the upstream stagnation point located at $x=y=z=0$, we may replace $r = R - x$, such that
\begin{align}
	u_x &= u_\infty \left( 1 - \frac{3}{2} \left( 1 - \frac{x}{R} \right)^{-1} + \frac{1}{2} \left( 1 - \frac{x}{R} \right)^{-3} \right)\\
 &= u_\infty \left( 1 - \frac{x}{R} \right)^{-3} \left( \frac{x}{R} \right)^2 \left( \frac{3}{2} - \frac{x}{R} \right)
 = u_\infty g \left(  - \frac{x}{R} \right),
\end{align}
where $g(z)$ is given in \cref{eq:gzdef}.

We now seek the leading-order contribution away from the separatrix.
Close to the surface, we can expand the field \eqref{eq:u_rtheta_sphere} in the small distance $\delta$ defined by
\begin{align}
    r = R - \delta .
    \label{eq:def_r_R_delta}
\end{align}
To the 2nd order in $\delta$,
\begin{subequations}
\begin{align}
    u_r &= \frac{3 u_\infty}{2} \delta^2 \cos \theta, \\
    u_\theta &= - \frac{3 u_\infty}{2} \delta \left( 1 - \frac{3}{2} \delta \right) \sin \theta.
\end{align}\label{eq:u_rtheta_sphere_expanded}\end{subequations}
Close to the stagnation point, i.e.\ where $|x|,|y|,|z| \ll R$, we can expand $r = \sqrt{(x-R)^2 + y^2 + z^2}$ and combine with \cref{eq:def_r_R_delta} to obtain
\begin{align}
    \delta = - \frac{x}{R} + O(x^2, y^2, z^2).
\end{align}
Thus, to the leading order, we obtain from \cref{eq:u_rtheta_sphere_expanded} the velocity field in Cartesian coordinates:
\begin{align}
	\v u = \begin{bmatrix}
            \frac{3 u_\infty}{2 R^2} x^2 \\
            -\frac{3 u_\infty }{2 R^2} x y \\
            -\frac{3 u_\infty }{2 R^2} x z
	\end{bmatrix}.
\end{align}
We see that sufficiently close to the stagnation point, the flow field is consistent with the infinite plate expression \eqref{eq:u_infplate_3d} with $\alpha = 3 u_\infty / (2 R^2)$ and $\beta = 1/2$.

\section{Asymptotic behaviour by eigenfunction expansion}
\label{sec:app:finiteD}

Here, we seek the asymptotic behaviour of a strip/sheet compressed against a stagnation point.
We consider the one-dimensional advection-diffusion equation evaluated along the separatrix, neglecting diffusion along the strip/sheet.
For convenience of notation, we consider the positive domain $x \in [0, \infty)$, such that the flow is in the negative direction.
For a generic dimensionless function $g$ describing the velocity field, cf. \cref{eq:ux_g}, we can then express
\begin{align}
    \pdt c = u_\infty g \left( \frac{x}{\ell} \right) \partial_x c + D \partial_x^2 c,
    \label{eq:pdtc_finiteD}
\end{align}
with the boundary condition $\partial_x c = 0$ at $x=0$ and the initial condition $c(x, 0) = C \delta(x-x_0)$.
Introducing the rescaled time and space coordinates,  $\tilde t = t / t_B$ and $\tilde x = x / s_B$, where $t_B, s_B$ are given in \cref{eq:stB_gen}, we obtain
\begin{align}
    \partial_{\tilde t} c = \frac{u_\infty t_B}{s_B} g\left( \frac{s_B}{\ell} \tilde x \right) \partial_{\tilde x} c + \frac{D t_B}{s_B^2}\partial_{\tilde x}^2 c
    = \kappa^{-1/3} \Peclet^{2/3} g\left( \kappa^{-1/3} \Peclet^{-1/3} \tilde x \right) \partial_{\tilde x} c + \partial_{\tilde x}^2 c
\end{align}
Skipping the tildes for brevity, we obtain the nondimensionalized expression
\begin{align}
    \pdt c &=  \mathcal{L} c, \\
    \textrm{where} \quad \mathcal{L} &= \kappa^{-1/3} \Peclet^{2/3} g\left( \kappa^{-1/3} \Peclet^{-1/3} x \right) \partial_{x} + \partial_{x}^2 .
    \label{eq:Loperator_def}
\end{align}
By separation of variables we can expand $c(x, t)$ as
\begin{align}
    c(x, t) = \sum_{i=0}^\infty c_i \phi_i(x) \exp(-\lambda_i t),
    \label{eq:expansion}
\end{align}
where the eigenfunctions $\phi_i(x)$ and eigenvalues $\lambda_i$ of the linear operator $\mathcal{L}$ are defined by
\begin{subequations}
\begin{align}
    \left( \lambda_i + \mathcal{L} \right) \phi_i &= 0, \quad i = 0, 1, 2, 3, \ldots, \label{eq:eigenvalue_problem} \\
    \partial_x \phi_i &= 0, \quad x=0, \\
    \phi_i &= 0, \quad x \to \infty.
\end{align}    
\end{subequations}
The set $\{ \phi_i \}$ form an orthonormal basis, i.e.\ $\int_0^{\infty} \phi_i (x) \phi_j (x) \, \diff x = \delta_{ij}$.
The expansion coefficients are found by multiplying \cref{eq:expansion} at $t=0$ by $\phi_j$ and integrating over the domain:
\begin{align}
    c_j = \int_0^\infty c(x, 0) \phi_j (x) \, \diff x = \int_0^\infty c_{\rm init} (x) \phi_j (x) \, \diff x = C \int_0^\infty \delta(x - x_0) \phi_j (x) \, \diff x = C \phi_j (x_0),
\end{align}
where we approximate the initial condition \cref{eq:c_0_x} with a Dirac delta distribution, with $C = \sqrt{2\pi} c_0 s_0$ (the length scale $s_0$ is also expressed in terms of $s_B$).
At large times, the smallest positive eigenvalue, which for simplicity we refer to as $\lambda$, will dominate, such that
\begin{align}
    c(x, t) = C \phi(x_0) \phi(x) \exp(-\lambda t),
\end{align}
where $\phi(x)$ is the eigenfunction corresponding to $\lambda$.
Using the fact that $\phi$ has its maximum at $x=0$, the asymptotic maximum concentration is then given by
\begin{align}
    c_{\rm max}(t) \to C \phi(x_0) \phi(0) \exp(-\lambda t).
\end{align}

We define the $n$th moment by:
\begin{align}
    M_n(t) = \int_0^\infty x^n c(x, t) \, \diff x \to \Phi_n e^{-\lambda t}, \quad \text{with} \quad \Phi_n = \int_0^\infty x^n \phi (x)  \, \diff x,
    \label{eq:moments}
\end{align}
where the limit is valid as $t\to\infty$.
The asymptotic mean position and the half-width are given by
\begin{align}
    \mean{x}_\infty = \lim_{t \to \infty} \frac{M_1(t)}{M_0(t)} = \frac{\Phi_{1}}{\Phi_{0}}, \quad
    \sigma_\infty = \lim_{t \to \infty} \sqrt{ \frac{M_2(t)}{M_0(t)} - \frac{M^2_1(t)}{M^2_0(t)}} = \sqrt{\frac{\Phi_{2}}{\Phi_{0}} - \frac{\Phi_{1}^2}{\Phi_{0}^2}}.
    \label{eq:xmeansigmainfdef}
\end{align}
This problem can be solved numerically for any given $\kappa$ and $\Peclet$ for general velocity fields to obtain the dependence of $\mean{x}_\infty$, $\sigma_\infty$ and $\lambda$ on $\Peclet$ for general velocity fields $g$.

\subsection{Infinite plate}
\label{sec:app:finiteD_infplate}
For the case of the infinite plate, $g(z) = \kappa z^2$ (using \cref{eq:alpha_def} and $z = x/\ell$) and \cref{eq:eigenvalue_problem} becomes
\begin{align}
    \left[ \lambda^{(0)} + \mathcal{L}^{(0)} \right] \phi^{(0)} = 0, \label{eq:expansion_eps0_infplate}
\end{align}
where $\lambda^{(0)}$ is the smallest positive eigenvalue and $\phi^{(0)}$ the corresponding eigenfunction of the operator
\begin{align}
    \mathcal{L}^{(0)} = x^2 \partial_x + \partial^2_x.
\end{align}
The superscript `(0)' is chosen because the problem is equivalent to a zeroth order expansion of a general velocity field $g$, as shown in \cref{sec:app:finiteD_pert} below.
The problem \eqref{eq:expansion_eps0_infplate} can be solved numerically and yields 
\begin{align}
    \lambda^{(0)} = 1.241, \quad
    \mean{x}_\infty^{(0)} = 0.447, \quad
    \sigma_\infty^{(0)} = 0.620.
    \label{eq:asymptotic_values_infplate}
\end{align}

\subsection{Perturbative solution}
\label{sec:app:finiteD_pert}
To obtain the asymptotic scaling of $\mean{x}_\infty(\Peclet)$, $\sigma_\infty (\Peclet)$ and $\lambda(\Peclet)$ beyond the idealized limit of an infinite plate, we may use the general expansion of $g(z)$ for small $z$ (see \cref{{eq:ux_general_exp}}),
\begin{align}
    g(z) = \frac{1}{2} g''(0) z^2 + \frac{1}{6} g'''(0) z^3 + O(z^4) = \kappa z^2 + \omega z^3 + O(z^4),
\end{align}
where we have implicitly introduced $\omega$.
Plugging the latter into \cref{eq:Loperator_def},
\begin{align}
    \mathcal{L} &=  \left[  x^2  + \omega \kappa^{-4/3} \Peclet^{-1/3} x^3 \right] \partial_{x} + \partial_{x}^2  + O(x^4)
    =   x^2 \left[ 1 + \epsilon x \right] \partial_{x} + \partial_{x}^2 + O(\epsilon^2),
\end{align}
where $\epsilon = \omega \kappa^{-4/3} \Peclet^{-1/3}$ can be considered a small parameter for large $\Peclet$.
We can then use perturbation theory to approximate the eigenvalues and eigenfunctions for small $\epsilon$ \cite[sec. 7.3]{bender2013advanced}.
Letting the superscripts denote the order of the expansion, we define
\begin{align}
    \phi &= \phi^{(0)} + \epsilon \phi^{(1)} + O(\epsilon^2), \\
    \lambda &= \lambda^{(0)} + \epsilon \lambda^{(1)} + O(\epsilon^2),
\end{align}
which becomes, using \cref{eq:eigenvalue_problem}, and collecting terms of equal order up to the first order in $\epsilon$:
\begin{align}
    \epsilon^0: &\quad \left[ \lambda^{(0)} + \mathcal{L}^{(0)} \right] \phi^{(0)} = 0, \label{eq:expansion_eps0} \\
    \epsilon^1: &\quad \left[ \lambda^{(0)} + \mathcal{L}^{(0)} \right] \phi^{(1)} = - x^3 \d{\phi^{(0)}}{x} - \lambda^{(1)} \phi^{(0)}. \label{eq:expansion_eps1}
\end{align}
As expected, the zeroth order problem \eqref{eq:expansion_eps0} is identical to the infinite plate case \eqref{eq:expansion_eps0_infplate}, and hence the zeroth-order asymptotic values are given by \cref{eq:asymptotic_values_infplate}.

To obtain the first-order correction, we make the ansatz $\phi^{(1)}(x) = \phi^{(0)}(x) f(x)$, where $f$ is an undetermined function.
Using the derivatives
\begin{align}
    \d {\phi^{(1)}}{x}  = \d{\phi^{(0)}}{x} f + \phi^{(0)} \d f x, \quad
    \dd {\phi^{(1)}}{x} = \dd{\phi^{(0)}}{x} f + 2 \d{\phi^{(0)}}{x} f' + \phi^{(0)} f'',
\end{align}
where $f' = \dinl{f}{x}$,
\cref{eq:expansion_eps1} becomes
\begin{align}
    f'' + (x^2 + 2 (\ln \phi^{(0)})') f' = - x^3 (\ln \phi^{(0)})' - \lambda^{(1)}. 
\end{align}
This yields
\begin{align}
    \d{}{x} \left[ \left(\phi^{(0)} \right)^2 \exp \left( \frac{x^3}{3} \right) f' \right] = - \phi^{(0)} \exp \left( \frac{x^3}{3} \right) \left[ x^3 \d{\phi^{(0)}}{x} + \lambda^{(1)} \phi^{(0)} \right] .
    \label{eq:expansion_eps1_ode}
\end{align}
Integrating from $x=0$ to $\infty$, and using $f'(0) = f'(\infty) \to 0$, we find
\begin{align}
     \lambda^{(1)} = -\frac{\int_0^\infty x^3 \phi^{(0)} \d{\phi^{(0)}}{x} \exp \left( \frac{x^3}{3} \right) \, \diff x }{ \int_0^\infty \left( \phi^{(0)} \right)^2 \exp \left( \frac{x^3}{3} \right) \, \diff x }
     = \frac{ \int_0^\infty (3 x^2 + x^5) Q(x) \, \diff x }{ 2 \int_0^\infty Q(x) \, \diff x } ,
     \label{eq:lambda_1}
\end{align}
where we have defined
\begin{align}
    Q(x) = \left( \phi^{(0)} (x) \right)^2 \exp \left( \frac{x^3}{3} \right)
\end{align}
We can integrate \cref{eq:expansion_eps1_ode} to obtain
\begin{align}
    f'(x) &= 
    - Q^{-1}(x) \left[ \int_0^x \phi^{(0)} (x') \exp \left( \frac{x'^3}{3} \right) x'^3 (\phi^{(0)})'(x') \, \diff x' + \lambda^{(1)} \int_0^x Q(x') \, \diff x' \right] \\
    &= - \frac{1}{2} x^3 + \frac{1}{2} Q^{-1}(x) \int_0^x \left(3 x'^2 + x'^5 \right) Q(x')  \, \diff x' - \lambda^{(1)} Q^{-1}(x) \int_0^x Q(x') \, \diff x' .
\end{align}
Defining
\begin{align}
    F(x) &= \int_0^x f'(x'') \, \diff x'' \\
    &= - \frac{1}{8} x^4 + \frac{1}{2} \int_0^x Q^{-1}(x'') \int_0^{x''} \left(3 x'^2 + x'^5 \right) Q(x') \, \diff x' \diff x'' - \lambda^{(1)} \int_0^x Q^{-1}(x'') \int_0^{x''} Q(x') \, \diff x' \diff x'',
    \label{eq:F_integral}
\end{align}
we can express
\begin{align}
     \phi^{(1)}(x) = \phi^{(0)} (x) \left( F(x) + f_0 \right),
     \label{eq:phi1_final}
\end{align}
where $f_0$ is an integration constant.
The latter is determined by requiring that the perturbed eigenfunction be normalized, i.e.\ 
\begin{align}
    \int_0^{\infty} \left( \phi^{(0)} + \epsilon \phi^{(1)} \right)^2 \, \diff x
    = \int_0^{\infty} \left( \phi^{(0)} (x) \right)^2 \, \diff x + 2 \epsilon \int_0^\infty \phi^{(0)} (x) \phi^{(1)} (x) \, \diff x + O(\epsilon^2) = 1,
\end{align}
that is, using $\int_0^\infty \left( \phi^{(0)} (x) \right)^2 \, \diff x = 1$,
we have 
    $\int_0^\infty \phi^{(0)} (x) \phi^{(1)} (x) \, \diff x = 0,$
which combined with \eqref{eq:phi1_final} yields
\begin{align}
     f_0 = - \int_0^\infty \left( \phi^{(0)} (x) \right)^2 F(x) \, \diff x .
     \label{eq:f0_integral}
\end{align}
We can now solve the integrals \eqref{eq:lambda_1}, \eqref{eq:f0_integral} and \eqref{eq:F_integral} numerically to obtain $\phi^{(1)}(x)$ through \cref{eq:phi1_final}.

Having obtained the eigenfunction describing the first-order corrections to the asymptotic behaviour, we now seek to quantify this correction in terms of $\mean{x}_\infty$ and $\sigma_\infty$.
Following \cref{eq:moments}, we express
\begin{align}
    \Phi_n^{(k)} = \int_0^\infty x^n \phi^{(k)} (x) \, \diff x,
    \quad \text{such that} \quad
    \Phi_n = \Phi_n^{(0)} + \epsilon \Phi_n^{(1)} + O(\epsilon^2).
\end{align}
Now, expanding \cref{eq:xmeansigmainfdef} to the first order in $\epsilon$, the mean position and the half-width are given respectively by
\begin{subequations}
    \begin{align}
    \mean{x}_\infty & 
    = \mean{x}_\infty^{(0)} + \epsilon \mean{x}_\infty^{(1)} + O(\epsilon^2), \quad \text{and} \quad
    \sigma_\infty 
    = \sigma_\infty^{(0)} + \epsilon \sigma_\infty^{(1)} + O(\epsilon^2),
\end{align}
where
\begin{align}
    \mean{x}_\infty^{(1)} &= \mean{x}_\infty^{(0)} \left[ \frac{\Phi_{1}^{(1)}}{\Phi_{1}^{(0)}} - \frac{\Phi_{0}^{(1)}}{\Phi_{0}^{(0)}} \right], \label{eq:xmean_inf_1} \\ 
    \sigma_\infty^{(1)} &= \frac{\sigma_\infty^{(0)}}{2} \left\{ \frac{\Phi_{2}^{(1)}}{\Phi_{2}^{(0)}} - \frac{\Phi_{0}^{(1)}}{\Phi_{0}^{(0)}} + \left( \frac{\mean{x}_\infty^{(0)}}{\sigma_\infty^{(0)}}\right)^2 \left[ \frac{\Phi_{2}^{(1)}}{\Phi_{2}^{(0)}} - 2 \frac{\Phi_{1}^{(1)}}{\Phi_{1}^{(0)}} + \frac{\Phi_{0}^{(1)}}{\Phi_{0}^{(0)}} \right] \right\}. \label{eq:sigma_inf_1}
\end{align}
\end{subequations}
Solving \cref{eq:lambda_1,eq:F_integral,eq:f0_integral,eq:phi1_final} numerically and using \cref{eq:xmean_inf_1,eq:sigma_inf_1,eq:asymptotic_values_infplate} we find the constants
\begin{align}
    \lambda^{(1)} = 1.07, \quad
    \mean{x}_\infty^{(1)} = -0.34, \quad
    \sigma_\infty^{(1)} = -0.27.
    \label{eq:asymptotic_values_1}
\end{align}
Reverting to physical units, inserting the definition of $\epsilon$, we end up with \cref{eq:asymptotic_scales_1}.

\end{appen}

\bibliographystyle{jfm}
\bibliography{references}

\newcommand{\acisc}{Adv. Colloid Interface Sci.} \newcommand{\ACMTOMS}{ACM
  Trans. Math. Softw.} \newcommand{\AdvGeophys}{Adv. Geophys.}
  \newcommand{\AdvPhys}{Adv. Phys.} \newcommand{\AdvWatRes}{Adv. Water Resour.}
  \newcommand{\AIPConfProc}{AIP Conf. Proc.} \newcommand{\AnalChem}{Anal.
  Chem.} \newcommand{\AnnPhys}{Ann. Phys.} \newcommand{\AnnuRevCon}{Annu. Rev.
  Control} \newcommand{\ApplMechRev}{Appl. Mech. Rev.}
  \newcommand{\ApplPhysLett}{Appl. Phys. Lett.} \newcommand{\ARFM}{Annu. Rev.
  Fluid Mech.} \newcommand{\BrazJPhys}{Braz. J. Phys.}
  \newcommand{\BrHeartJ}{Br. Heart J.} \newcommand{\CommunComputPhys}{Commun.
  Comput. Phys.} \newcommand{\CRAcadSci}{C. R. Acad. Sci.}
  \newcommand{\CRAcadSciIIB}{C. R. Acad. Sci. IIB} \newcommand{\CRMecanique}{C.
  R. M{'e}canique} \newcommand{\ComputFluids}{Comput. Fluids}
  \newcommand{\ComputMethodsApplMechEng}{Comput. Methods Appl. Mech. Eng.}
  \newcommand{\cpc}{Comput. Phys. Commun.} \newcommand{\csurfa}{Colloids Surf.
  A} \newcommand{\EPSL}{Earth Planet. Sci. Lett.}
  \newcommand{\EarthSciRev}{Earth-Sci. Rev.} \newcommand{\ejam}{Eur. J. Appl.
  Math.} \newcommand{\EnergyFuels}{Energy Fuels}
  \newcommand{\EnvironSciTechnol}{Environ. Sci. Technol.}
  \newcommand{\EPL}{Europhys. Lett.}
  \newcommand{\ESAIMMathModelNumerAnal}{ESAIM Math. Model. Numer. Anal.}
  \newcommand{\EurJMB}{Eur. J. Mech. B Fluids} \newcommand{\EurPhysJH}{Eur.
  Phys. J. H} \newcommand{\FaradayDiscuss}{Faraday Discuss.}
  \newcommand{\gji}{Geophys. J. Int.} \newcommand{\grl}{Geophys. Res. Lett.}
  \newcommand{\ieeetps}{IEEE Trans. Plasma Sci.} \newcommand{\IJGGC}{Int. J.
  Greenh. Gas Con.} \newcommand{\ijhmt}{Int. J. Heat Mass Transf.}
  \newcommand{\ijhff}{Int. J. Heat Fluid Fl.} \newcommand{\ijms}{Int. J. Mol.
  Sci.} \newcommand{\ijmf}{Int. J. Multiph. Flow} \newcommand{\ijnmf}{Int. J.
  Numer. Methods Fluids} \newcommand{\IJRMMS}{Int. J. Rock Mech. Min. Sci.}
  \newcommand{\InterfaceFreeBound}{Interface Free Bound.}
  \newcommand{\JAcoustSocAm}{J. Acoust. Soc. Am.} \newcommand{\JAmChemSoc}{J.
  Am. Chem. Soc.} \newcommand{\JApplPhys}{J. Appl. Phys.} \newcommand{\jast}{J.
  Adhes. Sci. Technol.} \newcommand{\JChemPhys}{J. Chem. Phys.}
  \newcommand{\jcis}{J. Colloid Interface Sci.} \newcommand{\jcompm}{J. Comp.
  Math.} \newcommand{\jcompp}{J. Comput. Phys.} \newcommand{\jcontamhydrol}{J.
  Contam. Hydrol.} \newcommand{\jfm}{J. Fluid Mech.} \newcommand{\jgrse}{J.
  Geophys. Res. Solid Earth} \newcommand{\jheattransf}{J. Heat Transf.}
  \newcommand{\JHydrol}{J. Hydrol.} \newcommand{\jmems}{J. Microelectromech.
  Syst.} \newcommand{\jpc}{J. Phys. Chem.} \newcommand{\JPhysChemB}{J. Phys.
  Chem. B} \newcommand{\jpcm}{J. Phys. Condens. Matter}
  \newcommand{\JPetSciEng}{J. Pet. Sci. Eng.} \newcommand{\JPSciEng}{J. Pet.
  Sci. Eng.} \newcommand{\JStatPhys}{J. Stat. Phys.} \newcommand{\labchip}{Lab
  Chip} \newcommand{\MarPetGeol}{Mar. Pet. Geol.} \newcommand{\MathComp}{Math.
  Comp.} \newcommand{\MathProcCambPhilosSoc}{Math. Proc. Camb. Philos. Soc.}
  \newcommand{\mmmas}{Math. Models Methods Appl. Sci.}
  \newcommand{\natgeo}{Nat. Geosci.} \newcommand{\natphys}{Nat. Phys.}
  \newcommand{\nedes}{Nuclear Engineering and Design}
  \newcommand{\NumerMath}{Numer. Math.} \newcommand{\pageoph}{Pure. Appl.
  Geophys.} \newcommand{\PhilosMag}{Philos. Mag.}
  \newcommand{\physicaa}{Physica A} \newcommand{\physicad}{Physica D}
  \newcommand{\physfluids}{Phys. Fluids} \newcommand{\physfluidsa}{Phys. Fluids
  A} \newcommand{\PhysRep}{Phys. Rep.} \newcommand{\PhysRev}{Phys. Rev.}
  \newcommand{\physreve}{Phys. Rev. E} \newcommand{\PhysRevFluids}{Phys. Rev.
  Fluids} \newcommand{\physrevlett}{Phys. Rev. Lett.} \newcommand{\PhysZ}{Phyz.
  Z.} \newcommand{\PNAS}{Proc. Natl. Acad. Sci.} \newcommand{\prsa}{Proc. Royal
  Soc. A} \newcommand{\ptrsl}{Philos. Trans. R. Soc.}
  \newcommand{\ptrsla}{Philos. Trans. R. Soc. A} \newcommand{\RepProgPhys}{Rep.
  Prog. Phys.} \newcommand{\RevGeophys}{Rev. Geophys.}
  \newcommand{\RevMineralGeochem}{Rev. Mineral. Geochem.}
  \newcommand{\revmodphys}{Rev. Mod. Phys.} \newcommand{\SciRep}{Sci. Rep.}
  \newcommand{\SIAP}{SIAM J. Appl. Math.} \newcommand{\SINUM}{SIAM J. Numer.
  Anal.} \newcommand{\SISC}{SIAM J. Sci. Comput.} \newcommand{\SJSSC}{SIAM J.
  Sci. and Stat. Comput.} \newcommand{\transpporousmedia}{Transp. Porous Media}
  \newcommand{\waterresres}{Water Resour. Res.} \newcommand{\ZPhysB}{Z. Phys.
  B} \newcommand{\ZPhysChem}{Z. Phys. Chem.}
\begin{thebibliography}{61}
\expandafter\ifx\csname natexlab\endcsname\relax\def\natexlab#1{#1}\fi
\def\au#1{#1} \def\ed#1{#1} \def\yr#1{#1}\def\at#1{#1}\def\jt#1{\textit{#1}}
  \def\bt#1{#1}\def\bvol#1{\textbf{#1}} \def\vol#1{#1} \def\pg#1{#1}
  \def\publ#1{#1}\def\arxiv#1{#1}\def\org#1{#1}\def\st#1{\textit{#1}}

\bibitem[de~Anna {\em et~al.\/}(2014)de~Anna, Jimenez-Martinez, Tabuteau,
  Turuban, Le~Borgne, Derrien \& M{\'e}heust]{deanna2014mixing}
{\sc \au{de~Anna, P.}, \au{Jimenez-Martinez, J.}, \au{Tabuteau, H.},
  \au{Turuban, R.}, \au{Le~Borgne, T.}, \au{Derrien, M.} \& \au{M{\'e}heust,
  Y.}} \yr{2014}  \at{Mixing and reaction kinetics in porous media: An
  experimental pore scale quantification}.  \jt{\EnvironSciTechnol}
  \bvol{48}~(1),  \pg{508--516}.

\bibitem[de~Anna {\em et~al.\/}(2017)de~Anna, Quaife, Biros \&
  Juanes]{deanna2017prediction}
{\sc \au{de~Anna, P.}, \au{Quaife, B.}, \au{Biros, G.} \& \au{Juanes, R.}}
  \yr{2017}  \at{Prediction of the low-velocity distribution from the pore
  structure in simple porous media}.  \jt{\PhysRevFluids}  \bvol{2}~(12),
  \pg{124103}.

\bibitem[Aquino {\em et~al.\/}(2023)Aquino, Le~Borgne \&
  Heyman]{aquino2023fluid}
{\sc \au{Aquino, T.}, \au{Le~Borgne, T.} \& \au{Heyman, J.}} \yr{2023}
  \at{Fluid--solid reaction in porous media as a chaotic restart process}.
  \jt{\physrevlett}  \bvol{130}~(26),  \pg{264001}.

\bibitem[Batchelor(1959)]{batchelor1959small}
{\sc \au{Batchelor, G.~K.}} \yr{1959}  \at{Small-scale variation of convected
  quantities like temperature in turbulent fluid part 1. general discussion and
  the case of small conductivity}.  \jt{\jfm}  \bvol{5}~(1),  \pg{113--133}.

\bibitem[Batchelor(2000)]{batchelor2000introduction}
{\sc \au{Batchelor, G.~K.}} \yr{2000} {\em An Introduction to Fluid
  Dynamics\/}.  \publ{Cambridge: Cambridge University Press}.

\bibitem[Bender \& Orszag(2013)]{bender2013advanced}
{\sc \au{Bender, C.~M.} \& \au{Orszag, S.~A.}} \yr{2013} {\em Advanced
  mathematical methods for scientists and engineers I: Asymptotic methods and
  perturbation theory\/}.  \publ{Springer Science \& Business Media}.

\bibitem[Berkowitz \& Scher(1995)]{berkowitz1995characterization}
{\sc \au{Berkowitz, B.} \& \au{Scher, H.}} \yr{1995}  \at{On characterization
  of anomalous dispersion in porous and fractured media}.  \jt{\waterresres}
  \bvol{31}~(6),  \pg{1461--1466}.

\bibitem[Bijeljic {\em et~al.\/}(2011)Bijeljic, Mostaghimi \&
  Blunt]{bijeljic2011signature}
{\sc \au{Bijeljic, B.}, \au{Mostaghimi, P.} \& \au{Blunt, M.~J.}} \yr{2011}
  \at{Signature of non-fickian solute transport in complex heterogeneous porous
  media}.  \jt{\physrevlett}  \bvol{107}~(20),  \pg{204502}.

\bibitem[Borgman {\em et~al.\/}(2023)Borgman, Turuban, G{\'e}raud, Le~Borgne \&
  M{\'e}heust]{borgman2023solute}
{\sc \au{Borgman, O.}, \au{Turuban, R.}, \au{G{\'e}raud, B.}, \au{Le~Borgne,
  T.} \& \au{M{\'e}heust, Y.}} \yr{2023}  \at{Solute front shear and
  coalescence control concentration gradient dynamics in porous micromodel}.
  \jt{Geophysical Research Letters}  \bvol{50}~(5),  \pg{e2022GL101407}.

\bibitem[Chen {\em et~al.\/}(2024)Chen, Lee \& Kang]{chen2024inertia}
{\sc \au{Chen, M.~A.}, \au{Lee, S.~H.} \& \au{Kang, P.~K.}} \yr{2024}
  \at{Inertia-induced mixing and reaction maximization in laminar porous media
  flows}.  \jt{\PNAS}  \bvol{121}~(50),  \pg{e2407145121}.

\bibitem[Conti \& Van~Dyke(1969)]{conti1969inviscid}
{\sc \au{Conti, R.} \& \au{Van~Dyke, M.}} \yr{1969}  \at{Inviscid reacting flow
  near a stagnation point}.  \jt{\jfm}  \bvol{35}~(4),  \pg{799--813}.

\bibitem[Dentz {\em et~al.\/}(2022)Dentz, Hidalgo \& Lester]{dentz2022mixing}
{\sc \au{Dentz, M.}, \au{Hidalgo, J.~J.} \& \au{Lester, D.}} \yr{2022}
  \at{Mixing in porous media: Concepts and approaches across scales}.
  \jt{\transpporousmedia}  \pg{pp. 1--49}.

\bibitem[Dentz {\em et~al.\/}(2011)Dentz, Le~Borgne, Englert \&
  Bijeljic]{dentz2011mixing}
{\sc \au{Dentz, M.}, \au{Le~Borgne, T.}, \au{Englert, A.} \& \au{Bijeljic, B.}}
  \yr{2011}  \at{Mixing, spreading and reaction in heterogeneous media: A brief
  review}.  \jt{\jcontamhydrol}  \bvol{120},  \pg{1--17}.

\bibitem[Dentz {\em et~al.\/}(2016)Dentz, Lester, Le~Borgne \&
  De~Barros]{dentz2016coupled}
{\sc \au{Dentz, M.}, \au{Lester, D.~R.}, \au{Le~Borgne, T.} \& \au{De~Barros,
  F.~P.}} \yr{2016}  \at{Coupled continuous-time random walks for fluid
  stretching in two-dimensional heterogeneous media}.  \jt{\physreve}
  \bvol{94}~(6),  \pg{061102}.

\bibitem[Farhat {\em et~al.\/}(2025)Farhat, Bolster \&
  Sole-Mari]{farhat2025linking}
{\sc \au{Farhat, S.}, \au{Bolster, D.} \& \au{Sole-Mari, G.}} \yr{2025}
  \at{Linking mixing interface deformation to concentration gradients in porous
  media}.  \jt{\PhysRevFluids}  \bvol{10}~(2),  \pg{024501}.

\bibitem[Feder {\em et~al.\/}(2022)Feder, Flekk{\o}y \&
  Hansen]{feder2022physics}
{\sc \au{Feder, J.}, \au{Flekk{\o}y, E.~G.} \& \au{Hansen, A.}} \yr{2022} {\em
  Physics of flow in porous media\/}.  \publ{Cambridge University Press}.

\bibitem[Fiori {\em et~al.\/}(2025)Fiori, Pramanik \& MacMinn]{fiori2025solute}
{\sc \au{Fiori, M.}, \au{Pramanik, S.} \& \au{MacMinn, C.~W.}} \yr{2025}
  \at{Solute transport due to periodic loading in a soft porous material}.
  \jt{\jfm}  \bvol{1009},  \pg{A15}.

\bibitem[Flekk{\o}y {\em et~al.\/}(1995)Flekk{\o}y, Oxaal, Feder \&
  J{\o}ssang]{flekkoy1995hydrodynamic}
{\sc \au{Flekk{\o}y, E.~G.}, \au{Oxaal, U.}, \au{Feder, J.} \& \au{J{\o}ssang,
  T.}} \yr{1995}  \at{Hydrodynamic dispersion at stagnation points: Simulations
  and experiments}.  \jt{\physreve}  \bvol{52}~(5),  \pg{4952}.

\bibitem[Gouillart {\em et~al.\/}(2007)Gouillart, Kuncio, Dauchot, Dubrulle,
  Roux \& Thiffeault]{gouillart2007walls}
{\sc \au{Gouillart, E.}, \au{Kuncio, N.}, \au{Dauchot, O.}, \au{Dubrulle, B.},
  \au{Roux, S.} \& \au{Thiffeault, J.-L.}} \yr{2007}  \at{Walls inhibit chaotic
  mixing}.  \jt{\physrevlett}  \bvol{99}~(11),  \pg{114501}.

\bibitem[Hallack {\em et~al.\/}(2025)Hallack, Sole-Mari, Farhat \&
  Bolster]{hallack20253d}
{\sc \au{Hallack, D. M.~C.}, \au{Sole-Mari, G.}, \au{Farhat, S.} \&
  \au{Bolster, D.}} \yr{2025}  \at{3d pore-scale mixing interface evolution}.
  \jt{ARC Geophysical Research}  \bvol{2}.

\bibitem[Heyman {\em et~al.\/}(2021)Heyman, Lester \& Le~Borgne]{Heyman2021}
{\sc \au{Heyman, J.}, \au{Lester, D.~R.} \& \au{Le~Borgne, T.}} \yr{2021}
  \at{Scalar signatures of chaotic mixing in porous media}.  \jt{\physrevlett}
  \bvol{126},  \pg{034505}.

\bibitem[Heyman {\em et~al.\/}(2020)Heyman, Lester, Turuban, M{\'e}heust \&
  Le~Borgne]{Heyman2020}
{\sc \au{Heyman, J.}, \au{Lester, D.~R.}, \au{Turuban, R.}, \au{M{\'e}heust,
  Y.} \& \au{Le~Borgne, T.}} \yr{2020}  \at{Stretching and folding sustain
  microscale chemical gradients in porous media}.  \jt{\PNAS}  \bvol{117}~(24),
   \pg{13359--13365}.

\bibitem[Izumoto {\em et~al.\/}(2024)Izumoto, Escala, Heyman, Le~Borgne \&
  De~Wit]{izumoto2024control}
{\sc \au{Izumoto, S.}, \au{Escala, D.~M.}, \au{Heyman, J.}, \au{Le~Borgne, T.}
  \& \au{De~Wit, A.}} \yr{2024}  \at{Control of chemical waves by fluid
  stretching and compression}.  \jt{\physrevlett}  \bvol{133}~(21),
  \pg{218001}.

\bibitem[Izumoto {\em et~al.\/}(2023)Izumoto, Heyman, Huisman, De~Vriendt,
  Soulaine, Gomez, Tabuteau, M{\'e}heust \& Le~Borgne]{izumoto2023enhanced}
{\sc \au{Izumoto, S.}, \au{Heyman, J.}, \au{Huisman, J.~A.}, \au{De~Vriendt,
  K.}, \au{Soulaine, C.}, \au{Gomez, F.}, \au{Tabuteau, H.}, \au{M{\'e}heust,
  Y.} \& \au{Le~Borgne, T.}} \yr{2023}  \at{Enhanced mixing and reaction in
  converging flows: Theory and pore-scale imaging}.  \jt{\waterresres}
  \bvol{59}~(8),  \pg{e2023WR034749}.

\bibitem[Langlois \& Deville(2014)]{langlois2014slow}
{\sc \au{Langlois, W.~E.} \& \au{Deville, M.~O.}} \yr{2014} {\em Slow Viscous
  Flow\/}, 2nd edn.  \publ{Cham: Springer International Publishing}.

\bibitem[Le~Borgne {\em et~al.\/}(2015)Le~Borgne, Dentz \&
  Villermaux]{LeBorgne2015}
{\sc \au{Le~Borgne, T.}, \au{Dentz, M.} \& \au{Villermaux, E.}} \yr{2015}
  \at{The lamellar description of mixing in porous media}.  \jt{\jfm}
  \bvol{770},  \pg{458--498}.

\bibitem[Le~Borgne \& Heyman(2026)]{le2026fluid}
{\sc \au{Le~Borgne, T.} \& \au{Heyman, J.}} \yr{2026}  \at{Fluid deformation
  and mixing in porous media as drivers for chemical and biological processes}.
   \jt{\ARFM}  \bvol{58}.

\bibitem[Lester \& Dentz(2025)]{lester2025line}
{\sc \au{Lester, D.} \& \au{Dentz, M.}} \yr{2025}  \at{Line stretching in
  random flows}.  \jt{arXiv preprint arXiv:2504.17982} .

\bibitem[Lester {\em et~al.\/}(2021)Lester, Dentz, Bandopadhyay \&
  Le~Borgne]{lester2021lagrangian}
{\sc \au{Lester, D.~R.}, \au{Dentz, M.}, \au{Bandopadhyay, A.} \&
  \au{Le~Borgne, T.}} \yr{2021}  \at{The {L}agrangian kinematics of
  three-dimensional {D}arcy flow}.  \jt{\jfm}  \bvol{918},  \pg{A27}.

\bibitem[Lester {\em et~al.\/}(2016)Lester, Dentz \&
  Le~Borgne]{lester2016chaotic}
{\sc \au{Lester, D.~R.}, \au{Dentz, M.} \& \au{Le~Borgne, T.}} \yr{2016}
  \at{Chaotic mixing in three-dimensional porous media}.  \jt{\jfm}
  \bvol{803},  \pg{144--174}.

\bibitem[Lester {\em et~al.\/}(2018)Lester, Dentz, Le~Borgne \&
  de~Barros]{lester2018fluid}
{\sc \au{Lester, D.~R.}, \au{Dentz, M.}, \au{Le~Borgne, T.} \& \au{de~Barros,
  F.~P.}} \yr{2018}  \at{Fluid deformation in random steady three-dimensional
  flow}.  \jt{\jfm}  \bvol{855},  \pg{770--803}.

\bibitem[Lester {\em et~al.\/}(2025)Lester, Heyman, Méheust \&
  Le~Borgne]{lester2025unified}
{\sc \au{Lester, D.~R.}, \au{Heyman, J.}, \au{Méheust, Y.} \& \au{Le~Borgne,
  T.}} \yr{2025}  \at{A unified theory of pore-scale chaotic advection}.
  \jt{\jfm}  \bvol{1017},  \pg{A13}.

\bibitem[Lester {\em et~al.\/}(2013)Lester, Metcalfe \& Trefry]{Lester2013is}
{\sc \au{Lester, D.~R.}, \au{Metcalfe, G.} \& \au{Trefry, M.}} \yr{2013}
  \at{Is chaotic advection inherent to porous media flow?}  \jt{\physrevlett}
  \bvol{111}~(17),  \pg{174101}.

\bibitem[Liao {\em et~al.\/}(2015)Liao, Dasgupta, Srinivasan \&
  Liu]{liao2015mixing}
{\sc \au{Liao, H.}, \au{Dasgupta, P.~K.}, \au{Srinivasan, K.} \& \au{Liu, Y.}}
  \yr{2015}  \at{Mixing characteristics of mixers in flow analysis. application
  to two-dimensional detection in ion chromatography}.  \jt{\AnalChem}
  \bvol{87}~(1),  \pg{793--800}.

\bibitem[Lienhard(2020)]{lienhard2020heat}
{\sc \au{Lienhard, J.~H.}} \yr{2020}  \at{Heat transfer in flat-plate boundary
  layers: A correlation for laminar, transitional, and turbulent flow}.
  \jt{\jheattransf}  \bvol{142}~(6),  \pg{061805}.

\bibitem[Linga(2026)]{linga2026partrac}
{\sc \au{Linga, G.}} \yr{2026} Partrac - particle tracking in time-dependent
  multiphase flow fields. Https://github.com/gautelinga/partrac.

\bibitem[Linga {\em et~al.\/}(2026)Linga, Pierce, Moura, Mathiesen, Renard \&
  Le~Borgne]{linga2026dynamic}
{\sc \au{Linga, G.}, \au{Pierce, K.}, \au{Moura, M.}, \au{Mathiesen, J.},
  \au{Renard, F.} \& \au{Le~Borgne, T.}} \yr{2026}  \at{Dynamic multiphase flow
  triggers chaotic mixing in porous media}.  \jt{arXiv preprint
  arXiv:2604.1038} .

\bibitem[Mart{\'\i}nez-Ruiz {\em et~al.\/}(2018)Mart{\'\i}nez-Ruiz, Meunier,
  Favier, Duchemin \& Villermaux]{martinez2018diffusive}
{\sc \au{Mart{\'\i}nez-Ruiz, D.}, \au{Meunier, P.}, \au{Favier, B.},
  \au{Duchemin, L.} \& \au{Villermaux, E.}} \yr{2018}  \at{The diffusive sheet
  method for scalar mixing}.  \jt{\jfm}  \bvol{837},  \pg{230--257}.

\bibitem[Mathiesen {\em et~al.\/}(2023)Mathiesen, Linga, Misztal, Renard \&
  Le~Borgne]{mathiesen2023dynamic}
{\sc \au{Mathiesen, J.}, \au{Linga, G.}, \au{Misztal, M.}, \au{Renard, F.} \&
  \au{Le~Borgne, T.}} \yr{2023}  \at{Dynamic fluid connectivity controls solute
  dispersion in multiphase porous media flow}.  \jt{\grl}  \bvol{50}~(16),
  \pg{e2023GL105233}.

\bibitem[Meunier \& Villermaux(2010)]{meunier2010diffusive}
{\sc \au{Meunier, P.} \& \au{Villermaux, E.}} \yr{2010}  \at{The diffusive
  strip method for scalar mixing in two dimensions}.  \jt{\jfm}  \bvol{662},
  \pg{134--172}.

\bibitem[Ottino(1989)]{ottino1989kinematics}
{\sc \au{Ottino, J.~M.}} \yr{1989} {\em The Kinematics of Mixing: Stretching,
  Chaos, and Transport\/}. {\em Cambridge Texts in Applied Mathematics\/} 3.
  \publ{Cambridge: Cambridge University Press}.

\bibitem[Oxaal {\em et~al.\/}(1994)Oxaal, Flekk{\o}y \&
  Feder]{oxaal1994irreversible}
{\sc \au{Oxaal, U.}, \au{Flekk{\o}y, E.~G.} \& \au{Feder, J.}} \yr{1994}
  \at{Irreversible dispersion at a stagnation point: Experiments and lattice
  {B}oltzmann simulations}.  \jt{\physrevlett}  \bvol{72}~(22),  \pg{3514}.

\bibitem[Pierce {\em et~al.\/}(2026)Pierce, Le~Borgne, Renard \&
  Linga]{pierce2026pore}
{\sc \au{Pierce, J.~K.}, \au{Le~Borgne, T.}, \au{Renard, F.} \& \au{Linga, G.}}
  \yr{2026}  \at{How pore-scale disorder controls fluid stretching in porous
  media}.  \jt{arXiv preprint arXiv:2604.02981} .

\bibitem[Puyguiraud {\em et~al.\/}(2021)Puyguiraud, Gouze \&
  Dentz]{puyguiraud2021pore}
{\sc \au{Puyguiraud, A.}, \au{Gouze, P.} \& \au{Dentz, M.}} \yr{2021}
  \at{Pore-scale mixing and the evolution of hydrodynamic dispersion in porous
  media}.  \jt{\physrevlett}  \bvol{126}~(16),  \pg{164501}.

\bibitem[Ranz(1979)]{ranz1979applications}
{\sc \au{Ranz, W.~E.}} \yr{1979}  \at{Applications of a stretch model to
  mixing, diffusion, and reaction in laminar and turbulent flows}.  \jt{AIChE
  J.}  \bvol{25}~(1),  \pg{41--47}.

\bibitem[Rolle \& Le~Borgne(2019)]{rolle2019mixing}
{\sc \au{Rolle, M.} \& \au{Le~Borgne, T.}} \yr{2019}  \at{Mixing and reactive
  fronts in the subsurface}.  \jt{\RevMineralGeochem}  \bvol{85}~(1),
  \pg{111--142}.

\bibitem[Sanquer {\em et~al.\/}(2024)Sanquer, Heyman, Hanna \&
  Le~Borgne]{sanquer2024microscale}
{\sc \au{Sanquer, H.}, \au{Heyman, J.}, \au{Hanna, K.} \& \au{Le~Borgne, T.}}
  \yr{2024}  \at{Microscale chaotic mixing as a driver for chemical reactions
  in porous media}.  \jt{\EnvironSciTechnol}  \bvol{58}~(20),  \pg{8899--8908}.

\bibitem[Shafabakhsh {\em et~al.\/}(2024)Shafabakhsh, Le~Borgne, Renard \&
  Linga]{shafabakhsh2024resolving}
{\sc \au{Shafabakhsh, P.}, \au{Le~Borgne, T.}, \au{Renard, F.} \& \au{Linga,
  G.}} \yr{2024}  \at{Resolving pore-scale concentration gradients for
  transverse mixing and reaction in porous media}.  \jt{\AdvWatRes}
  \bvol{192},  \pg{104791}.

\bibitem[Shah \& Sekulic(1998)]{shah1998heat}
{\sc \au{Shah, R.} \& \au{Sekulic, D.}} \yr{1998}  \at{Heat exchangers}.
  \jt{Handbook of heat transfer}  \bvol{3}.

\bibitem[Sole-Mari {\em et~al.\/}(2023)Sole-Mari, Bolster \&
  Fernandez-Garcia]{sole2023closer}
{\sc \au{Sole-Mari, G.}, \au{Bolster, D.} \& \au{Fernandez-Garcia, D.}}
  \yr{2023}  \at{A closer look: High-resolution pore-scale simulations of
  solute transport and mixing through porous media columns}.
  \jt{\transpporousmedia}  \bvol{146}~(1),  \pg{85--111}.

\bibitem[Souzy {\em et~al.\/}(2020)Souzy, Lhuissier, M{\'e}heust, Le~Borgne \&
  Metzger]{souzy2020velocity}
{\sc \au{Souzy, M.}, \au{Lhuissier, H.}, \au{M{\'e}heust, Y.}, \au{Le~Borgne,
  T.} \& \au{Metzger, B.}} \yr{2020}  \at{Velocity distributions, dispersion
  and stretching in three-dimensional porous media}.  \jt{\jfm}  \bvol{891},
  \pg{A16}.

\bibitem[Souzy {\em et~al.\/}(2018)Souzy, Zaier, Lhuissier, Le~Borgne \&
  Metzger]{souzy2018mixing}
{\sc \au{Souzy, M.}, \au{Zaier, I.}, \au{Lhuissier, H.}, \au{Le~Borgne, T.} \&
  \au{Metzger, B.}} \yr{2018}  \at{Mixing lamellae in a shear flow}.  \jt{\jfm}
   \bvol{838},  \pg{R3}.

\bibitem[Steefel {\em et~al.\/}(2005)Steefel, DePaolo \&
  Lichtner]{steefel2005reactive}
{\sc \au{Steefel, C.~I.}, \au{DePaolo, D.~J.} \& \au{Lichtner, P.~C.}}
  \yr{2005}  \at{Reactive transport modeling: An essential tool and a new
  research approach for the {E}arth sciences}.  \jt{\EPSL}  \bvol{240}~(3-4),
  \pg{539--558}.

\bibitem[Tartakovsky(2013)]{tartakovsky2013assessment}
{\sc \au{Tartakovsky, D.~M.}} \yr{2013}  \at{Assessment and management of risk
  in subsurface hydrology: A review and perspective}.  \jt{\AdvWatRes}
  \bvol{51},  \pg{247--260}.

\bibitem[Taylor-West \& Hogg(2025)]{taylor2025stagnation}
{\sc \au{Taylor-West, J.~J.} \& \au{Hogg, A.~J.}} \yr{2025}  \at{Stagnation
  point flow of a viscoplastic fluid}.  \jt{Journal of Non-Newtonian Fluid
  Mechanics}  \pg{p. 105445}.

\bibitem[Turuban {\em et~al.\/}(2019)Turuban, Lester, Heyman, Le~Borgne \&
  M{\'e}heust]{turuban2019chaotic}
{\sc \au{Turuban, R.}, \au{Lester, D.~R.}, \au{Heyman, J.}, \au{Le~Borgne, T.}
  \& \au{M{\'e}heust, Y.}} \yr{2019}  \at{Chaotic mixing in crystalline
  granular media}.  \jt{\jfm}  \bvol{871},  \pg{562--594}.

\bibitem[Valocchi {\em et~al.\/}(2019)Valocchi, Bolster \&
  Werth]{valocchi2019mixing}
{\sc \au{Valocchi, A.~J.}, \au{Bolster, D.} \& \au{Werth, C.~J.}} \yr{2019}
  \at{Mixing-limited reactions in porous media}.  \jt{\transpporousmedia}
  \bvol{130}~(1),  \pg{157--182}.

\bibitem[Vel{\'a}squez-Parra {\em et~al.\/}(2022)Vel{\'a}squez-Parra, Aquino,
  Willmann, M{\'e}heust, Le~Borgne \&
  Jim{\'e}nez-Mart{\'\i}nez]{velasquez2022sharp}
{\sc \au{Vel{\'a}squez-Parra, A.}, \au{Aquino, T.}, \au{Willmann, M.},
  \au{M{\'e}heust, Y.}, \au{Le~Borgne, T.} \& \au{Jim{\'e}nez-Mart{\'\i}nez,
  J.}} \yr{2022}  \at{Sharp transition to strongly anomalous transport in
  unsaturated porous media}.  \jt{\grl}  \bvol{49}~(3),  \pg{e2021GL096280}.

\bibitem[Villermaux(2012)]{villermaux2012mixing}
{\sc \au{Villermaux, E.}} \yr{2012}  \at{Mixing by porous media}.  \jt{Comptes
  Rendus M{\'e}canique}  \bvol{340}~(11-12),  \pg{933--943}.

\bibitem[Villermaux(2019)]{Villermaux2019}
{\sc \au{Villermaux, E.}} \yr{2019}  \at{Mixing versus stirring}.  \jt{\ARFM}
  \bvol{51}~(1),  \pg{245--273}.

\bibitem[Yoon {\em et~al.\/}(2021)Yoon, Dentz \& Kang]{yoon2021optimal}
{\sc \au{Yoon, S.}, \au{Dentz, M.} \& \au{Kang, P.~K.}} \yr{2021}  \at{Optimal
  fluid stretching for mixing-limited reactions in rough channel flows}.
  \jt{\jfm}  \bvol{916},  \pg{A45}.

\end{thebibliography}

\end{document}